\documentclass{article}
\usepackage{amsmath}
\usepackage{graphicx}
\usepackage{hyperref}
\usepackage{subfig}

\begin{document}
\title{How Reliable are University Rankings?}
\author{Ali Dasdan, Eric Van Lare, and Bosko Zivaljevic\\
  KD Consulting\\
  Saratoga, CA, USA\\
  alidasdan@gmail.com\\
}

\maketitle

\begin{abstract}
University or college rankings have almost become an industry of their
own, published by US News \& World Report (USNWR) and similar
organizations. Most of the rankings use a similar scheme: Rank
universities in decreasing score order, where each score is computed
using a set of attributes and their weights; the attributes can be
objective or subjective while the weights are always subjective. This
scheme is general enough to be applied to ranking objects other than
universities. As shown in the related work, these rankings have
important implications and also many issues. In this paper, we take a
fresh look at this ranking scheme using the public College dataset; we
both formally and experimentally show in multiple ways that this
ranking scheme is not reliable and cannot be trusted as authoritative
because it is too sensitive to weight changes and can easily be
gamed. For example, we show how to derive reasonable weights
programmatically to move multiple universities in our dataset to the
top rank; moreover, this task takes a few seconds for over 600
universities on a personal laptop. Our mathematical formulation,
methods, and results are applicable to ranking objects other than
universities too. We conclude by making the case that all the data and
methods used for rankings should be made open for validation and
repeatability.
\end{abstract}

\section{Introduction}

Rankings of higher education institutions (universities for short)
have almost become an industry of its own~\cite{KiEb2016}. Most of the
rankings use a similar methodology: Select a set of numeric attributes
and a numeric weight for each attribute, then compute a final numeric
score as the sum of the products of each attribute with its
weight. The weight of an attribute determines the amount of
contribution the attribute makes to the final score. For the final
ranking, the universities are ranked in their decreasing score order.

This generic ranking methodology is simple enough that it has been
applied to ranking all kinds of objects, from universities to
hospitals to cities to countries~\cite{Oe2008}, e.g., see
\cite{niche2020a,usnews2020b,usnews2020a} for the rankings
methodologies for objects other than universities: Places to live,
hospitals, and countries. As a result, even though our focus in this
study is universities, the findings are applicable to other areas
where this generic ranking methodology is used.

There are many questions that have been explored about this ranking
methodology in general and university rankings in particular. For
example, why a given attribute is selected, whether a given attribute
has the correct or most up-to-date value, why an attribute is weighted
more than another one, what impact the opinion-based attributes have
on the final score, whether or the impact of these rankings in student
choices is warranted, whether or not a university should incentivize
their admins to improve their rank with a given ranking, whether or
not a university games these rankings when they share their data,
etc. There is a rich body of related work exploring many of these
issues, as explored in our related work section. There is also an
international effort that has provided a set of principles and
requirements (called the Berlin Principles~\cite{Berlin2019}) to
improve rankings and the practical implementations of the generic
ranking methodology.

We start our study in \S~\ref{sec:exercise} by applying the steps of a
generic ranking methodology codified in ten steps in an OECD
handbook~\cite{Oe2008}, to a public dataset, called the College
dataset, about more than a 1,000 US higher education
institutions. Though the College dataset was created in the year 1995,
the resulting ranking shows good alignment with the most recent
rankings from the well-known ranking organizations.

These well-known ranking organizations are the US News \& World Report
(USNWR)~\cite{usnews2019a}, Quacquarelli Symonds (QS)~\cite{qs2019a},
Times Higher Education (THE)~\cite{the2019a}, and ShanghaiRanking
Consultancy (SC)~\cite{china2019a}. We present the methodologies of
and the most recent rankings from these organizations in
\S~\ref{sec:four-rankings}. We also present two recent rankings of
universities in the computer science field, created by two groups of
academicians in the computer science department of a few US
universities.

We give a comprehensive but not exhaustive review of the rankings
literature, including the related Economics literature, in
\S~\ref{sec:related}. We lean towards providing references to survey
or overview papers as well as entry points to subfields so that the
interested readers can go deeper if they so desire.

We follow this by introducing the mathematical formulation that
underlies the generic ranking methodology. The mathematical
formulation uses concepts from linear algebra and integer linear
programming (ILP).

The main part of this paper is presented in \S~\ref{sec:explore}. We
formulate and solve six problems in this section. The main vehicle we
use in our formulation of these problems is the integer linear
programming. The ILP programs for these problems take a few seconds to
both generate and run for over 600 universities on a personal laptop.

In the first problem in \S~\ref{sec:random}, we explore the existence
of different rankings using Monte Carlo (randomized) simulation. We
show that there are many possible rankings and somewhat naturally
there are more than one university that can attain the top rank.

In the second problem in \S~\ref{sec:feasible}, we do the same
exploration but using the ILP for optimality. We confirm and go beyond
the findings of the simulation: We verify that many universities can
be moved to the top rank.

These two problems may show existence using weights that may not be
appealing to a human judge, e.g., they may be widely different from
each other, possibly yielding more than deserved impact on the final
score. In the third problem in \S~\ref{sec:appealing}, we rectify this
situation and derive appealing weights. We replicate the findings of
the first two problems using these new weights too.

These three problems show that there are many universities that can
attain the top rank, though not every university can do so. Then a
natural next question is how to find the best rank that each
university can attain. We attack this question in the fourth problem
in \S~\ref{sec:best}.

The first four problems always involve attribute weights, as in the
generic ranking methodology. In the fifth problem in
\S~\ref{sec:kemeny}, we explore a way of generating rankings without
weights, which uses the Kemeny rule~\cite{DwKuNa2001}. This no-weights
ranking approach eliminates many of the issues associated with weight
selection.

In the sixth and the last problem in \S~\ref{sec:delta}, we show how a
given university can improve its rank in a weight-based ranking. We
show that drastic rank improvements are possible by few attribute
value changes.

In this paper {\em our main thesis} is that {\em university rankings}
as commonly done today and somehow proposed as unique with so much
fanfare {\em are actually not reliable and can even be easily
  gamed}. We believe our results via these six problems provide strong
support to this thesis. This thesis applies especially to the
universities at the top rank. We show in multiple ways that it is
relatively easy to move multiple university to the top rank in a given
ranking. As we mentioned above, our findings are applicable to areas
where objects other than universities are ranked. In
\S~\ref{sec:reco}, we provide a discussion of these points together
with a few recommendations.

One unfortunate aspect of the rankings from the four well-known
rankings organizations is that their datasets and software code used
for their rankings are not in the public domain for repeatability. We
wanted to change this so we have posted our datasets and the related
software code in a public code repository~\cite{Da2020}. We hope the
well-known rankings organizations too will soon share their latest
datasets and the related software code in the public domain.

\section{A Ranking Exercise Using a Public Dataset}
\label{sec:exercise}

Rankings of higher education institutions were first started by the
U.S. Bureau of Education in 1870 and have been done by multiple other
organizations since; however, it can be argued that the current
rankings industry has been ignited by the U.S. News and World Report's
first ``America's Best Colleges'' ranking in
1983~\cite{KiEb2016}. Today, there are many rankings around the world,
some of which are the well-known worldwide rankings and some are
country specific~\cite{all2019}. The four well-known rankings that we
will cover in \S~\ref{sec:four-rankings} are the US News \& World
Report (USNWR), Quacquarelli Symonds (QS), Times Higher Education
(THE), and ShanghaiRanking Consultancy (SC).

In the sequel, for simplicity, we will use the term ``university'' to
refer to a university, a college, or a higher education institution,
regardless of the university is private or public, American or
international, i.e., one that belongs to another country.

Rankings are also common in areas other than higher education. For
example, there is a huge literature in Economics and related fields on
comparing countries on over 100 different human progress measures, the
most famous of which is probably the Human Development Index of the
United Nations~\cite{Oe2008,Ya2014}. These rankings use a ranking
methodology very similar to the one used for university rankings; as a
result, many of the results from this literature, including the
pitfalls, are applicable to university rankings.

As comprehensively outlined in the OECD rankings
handbook~\cite{Oe2008}, a typical ranking methodology follows ten
steps from step 1 to step 10, after the selection of the input
dataset, which we will refer to as step 0. We will apply these ten
steps to a ranking exercise of our own with the College dataset. All
the experiments reported in this paper were done on this dataset.

\subsection{Step 0: Gathering the Input Data}

In general, there are three basic sources of the input data for
university ranking~\cite{UsSa2007}:
\begin{enumerate}
\item[a)] surveys of the opinions of various stakeholders such as
  university or high school administrators;
\item[b)] independent third parties such as government agencies; and
\item[c)] university sources.
\end{enumerate}
The last source is now standardized under the initiative called the
``Common Data Set Initiative''~\cite{Cds2020}. A simple web search
with a university name followed by the query ``Common Data Set'' will
return many links to these data sets from many universities. Note that
among these three data sources, the survey data is inherently
subjective while the other two are supposed to be objective, although
unfortunately some intentional alteration of data by university
sources have been observed~\cite{Ri2019}.

Our data source fits into category b above: The College
dataset~\cite{college1995} is part of the StatLib datasets archive,
hosted at the Carnegie Mellon University; it contains data about many
(1,329 to be exact) but not all American higher education
institutions. Its collection in 1995 was facilitated by the American
Statistical Association. The two data sources are Association of
American University Professors (AAUP) and US News \& World Report
(USNWR), which contribute 17 and 35 attributes, respectively, per
university. There are many attributes with missing values for multiple
universities. See \cite{college1995} for the meaning of each
attribute. All the attributes in this dataset are objective.

\begin{figure}[ht]
  \centering
  \includegraphics[scale=0.3]{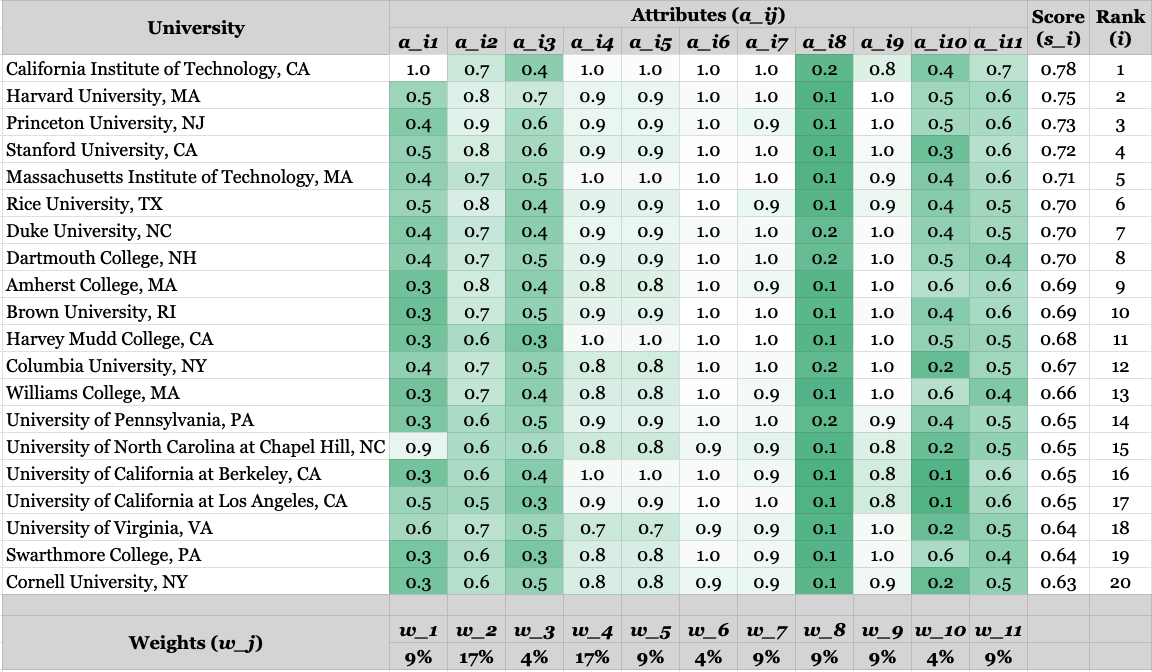}
  \caption{Our top 20 ranking generated from the College dataset. This
    ranking aligns well with the well-known rankings (partly by
    construction). Here the labels {\em a\_ij}, {\em w\_ij}, and {\em
      s\_i} refer to the attribute $a_{ij}$, weight $w_{j}$, and score
    $s_{i}$ of the university at rank $i$ for $j$ in 1 through
    11. Also here and in the sequel we use green color for
    highlighting attribute values.}
  \label{fig:top20ours}
\end{figure}

With the attributes and weights we selected, as detailed below, we
generated the top 20 universities as shown in
Fig.~\ref{fig:top20ours}. We hope the reader can appreciate that this
ranking is a reasonable one to the extent it aligns well with the
well-known rankings, to be discussed in a bit below.

One question that may come to the reader's mind could be the reason
for selecting this dataset and the relevance of this fairly old
dataset to the present. For the reason question, we wanted to make
sure that we choose a standard dataset that is available to all who
want to replicate our results; moreover, we do not have access to the
latest datasets used by the well-known rankings organizations. For the
relevance question, we ask the reader to review our mathematical and
problem formulations and convince themselves that our results are
applicable to any dataset containing a set of objects to rank using
their numerical attributes.

\subsection{Step 1: Developing a Theoretical or Conceptual Framework}

We have $n$ universities in some ranking, each of which has the same
$m$ attributes (also called variables or indicators in the Economics
literature) with potentially different values. We will use $i$ to
index universities and $j$ to index attributes. Each attribute
$a_{ij}$ of the $i$th university is associated with the same
real-number weight $w_{j}$. The score $s_{i}$ of the $i$th university
is a function of the attributes and weights of the university as
\begin{equation}
  s_{i} = g\left(\sum\limits_{j=1}^{m} w_{j} f\left( a_{ij} \right)\right),
\end{equation}
where the functions $g(\cdot)$ and $f(\cdot)$ usually reduce to the
identity function resulting in the following sum-of-products form:
\begin{equation}
  s_{i} = \sum\limits_{j=1}^{m} w_{j} a_{ij},
\end{equation}
where the $j$th weight determines the contribution of the $j$th
attribute to the final score.

A ranking of $n$ universities is a sorting of scores in decreasing
order such that the ``top'' ranked or the ``best'' university is the
university with the highest score or the one at rank 1. Later in
\S~\ref{sec:kemeny} we will discuss how to rank universities without
weights, in which case the ranking does not use scores.

Since this framework is well detailed in the mathematical formulation
section in \S~\ref{sec:math}, we will keep this section short. Also
see the weighting and aggregation section (\S~\ref{sec:aggr}) on how
the attributes and weights are manipulated for ranking.

\begin{figure}[ht]
  \centering
  \includegraphics[scale=0.3]{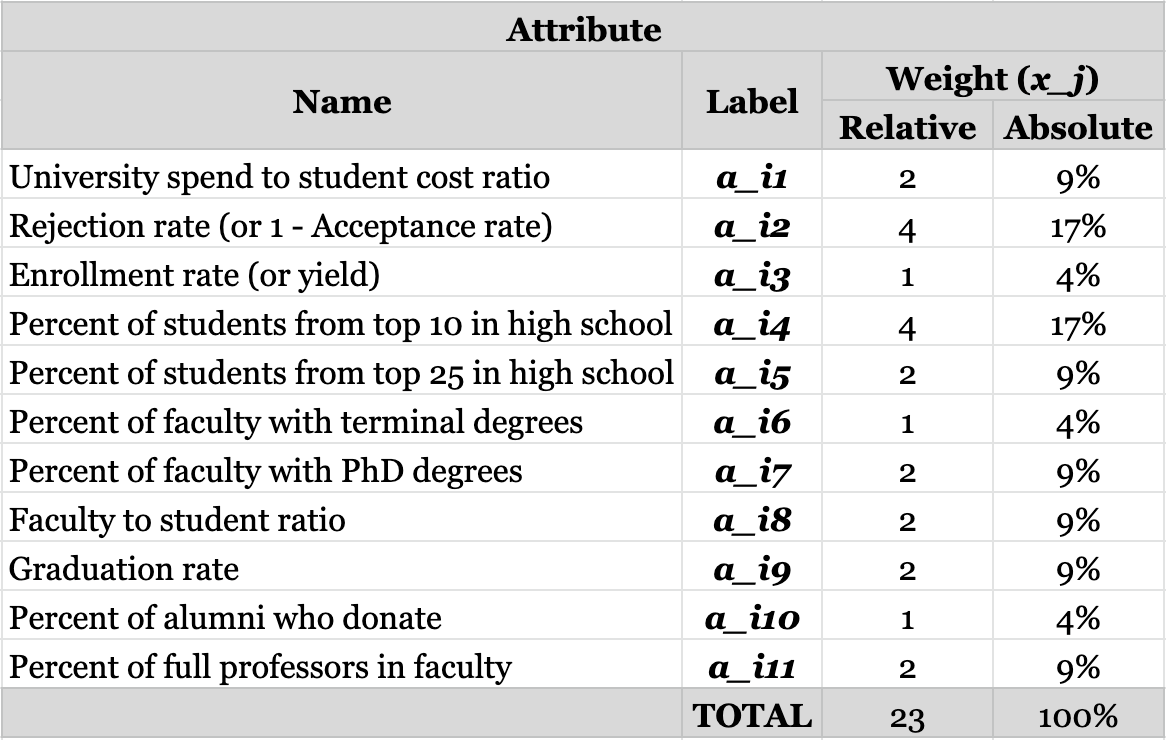}
  \caption{The 11 derived attributes and their assigned weights we
    used to rank the universities in the College dataset. Here the
    labels {\em a\_ij}, {\em w\_ij}, and {\em s\_i} refer to the
    attribute $a_{ij}$, weight $w_{j}$, and score $s_{i}$ of the
    university at rank $i$ for $j$ in 1 through 11.}
  \label{fig:attr}
\end{figure}

\subsection{Step 2: Selecting Attributes}
\label{sec:attr-select}

In general, the attributes for ranking or as indicators of quality in
higher education can be grouped into the following four
categories~\cite{FiUs2005}:
\begin{enumerate}
\item[a)] beginning characteristics, which cover the
  characteristics of the incoming students such as test scores or high
  school ranking;
\item[b)] learning inputs, which cover factors that help affect
  the learning experiences of students such as the financial resources
  of the university;
\item[c)] learning outputs, which cover the skill sets or any
  other attributes of graduates; and
\item[d)] final outcomes, which cover the outcomes of the
  educational system and what the students achieve in the end such as
  employment, income, and job satisfaction.
\end{enumerate}
It may be argued that category d is what matters most but most of the
current rankings focus on categories a and b because category d and to
some extent category c are difficult to measure continuously. Our
attributes due to what we can find in our dataset also fall into
categories a and b.

We prepared our dataset in two steps: 1) we joined the two source
datasets from AAUP and USNWR using the FICE code, a unique id per
university assigned by the American Federal Government. This generated
1,133 universities. 2) We selected 20 attributes out of the 52 total,
including the name and the state of the university. We then eliminated
any university with a missing value for any of the selected
attributes. This resulted in 603 universities.

The selected 20 attributes are: University, state, instructional
expenditure per student (1), in-state tuition (1), room and board
costs (1), room costs, board costs (1), additional fees (1), estimated
book costs (1), estimated personal spending (1), number of
applications received (2), number of applicants accepted (2), number
of new students enrolled (3), percent of new students from the top
10\% of their high school class (4), percent of new students from the
top 25\% of their high school class (5), percent of faculty with
terminal degree (6), percent of faculty with PhDs (5), student/faculty
ratio (8), graduation rate (9), percent of alumni who donate (10),
number of full professors (11), and number of faculty in all ranks
(11). The number $j$ in parenthesis indicates that the corresponding
attribute is used to derive the $j$th attribute in
Fig.~\ref{fig:attr}.

Why did we choose only 20 attributes out of the 52 total? Two reasons:
We wanted to use as many attributes as possible for each university;
we also wanted to make sure that the final list of attributes used in
ranking are comparable in the following senses:
\begin{enumerate}
  \item Every final attribute is either a percentage or a ratio. This
    ensures that they are comparable in magnitude.
  \item For every final attribute, a university with a higher value
    should be regarded by a prospective student as ``better'' than
    another university with a lower value.
\end{enumerate}
Taking these two into account, we could not select more than 20
initial attributes from the 52 total. We then converted these 20
attributes into a final list of 11 attributes (see
Fig.~\ref{fig:attr}), not counting the name and the state of the
university, using the following thought process that we think a
reasonable student would potentially go through:

``I, the student, want to go to a university $i$
\begin{enumerate}
\item[$a_{i1}$:] that spends for me far more than what it
  costs me in total (so that I get back more than what I put in);
\item[$a_{i2}$, $a_{i3}$:] that is desired highly by far more students
  that it can accommodate (so that I get a chance to study with top
  students);
\item[$a_{i4}$, $a_{i5}$:] that attracts the top students in their
  graduating class (so that I get a chance to study with top
  students);
\item[$a_{i6}$, $a_{i7}$:] that has more faculty with PhDs or other
  terminal degrees in their fields (so that I get taught by top
  researchers or teachers);
\item[$a_{i8}$:] that has a smaller student to faculty ratio (so that
  I can get more attention from professors);
\item[$a_{i9}$:] that has a higher graduation rate (so that I can
  graduate more easily);
\item[$a_{i10}$:] that has more of its alumni donating to the
  university (so that more financial resources are available to spend
  on students); and
\item[$a_{i11}$:] that has more of its classes taught by full
  professors (so that I get taught by top researchers or teachers).''
\end{enumerate}

Here the spend above indicates the total instructional expenditure by
the university per student and the cost to a student above covers the
tuition, room and board, fees, books, and personal expenses.

We hope the arguments made above look reasonable and we expect them to
be at least directionally correct. For example, a top researcher may
not be a top teacher but it feels reasonable to us to assume that with
a solid research experience there is some correlation towards better
qualifications to teach a particular subject.

The reader may or may not agree with this attribute selection process
but that is exactly one of the points of this paper: There is so much
subjectivity creeping in during multiple steps of the ranking
process. Later we will show how to remove some of this bias.

\subsection{Step 3: Imputation of Missing Data}

We then realized that we could easily repair some missing values: a)
if the value of the attribute ``room and board costs'' is missing, it
can easily be calculated by the sum of the values of the attributes
``room costs'' and ``board costs'' if both exist; b) University of
California campuses have very similar tuitions and fees, so we
substituted any missing value with the average of the remaining
values; c) For Stanford University, the values of the attributes
``additional fees'' and ``percent of faculty with PhDs'' were missing
so we replaced them with the values we were able to find or calculate
after some web searching. After these few repairs, we were able to
increase the number of universities very modestly, from 603 to 609 to
be exact, in our input dataset. We note that these repairs were not
necessary to reach our conclusions but if not done, well-regarded
universities like Stanford University and most of the University of
California campuses would have been missing from the final
ranking. 

\subsection{Step 4: Multivariate Analysis}

Using techniques such as principal component analysis or factor
analysis, this step looks at the underlying structure via the
independent components or factors of the attribute space. We will skip
this step for our dataset as this step is not highly relevant to the
theme of this paper. 

\subsection{Step 5: Normalization of Data}

Normalization ensures that each attribute falls into the same
interval, usually $[0, 1]$, in magnitude. This is required for one of
the ways of computing the final score for a university, the arithmetic
mean method~\cite{Oe2008,To2014}.

In our case, every attribute except for the ``university spend to
student cost ratio'' already falls in the unit interval $[0, 1]$. This
is because all our attributes are percentages or ratios by
construction. To minimize the impact of normalization on ranking, we
decided to normalize the only exception (i.e., the university spend to
student cost ratio) using the following formula (called the min-max
normalization~\cite{Oe2008}):
\begin{equation}
  a_{new} = \frac{a_{old} - min_{a}}{max_{a}-min_{a}}, 
\end{equation}
where $max_{a}$ and $min_{a}$ are the maximum and minimum values for
the attribute in question among the selected universities.

\begin{figure}[ht]
  \centering
  \includegraphics[scale=0.3]{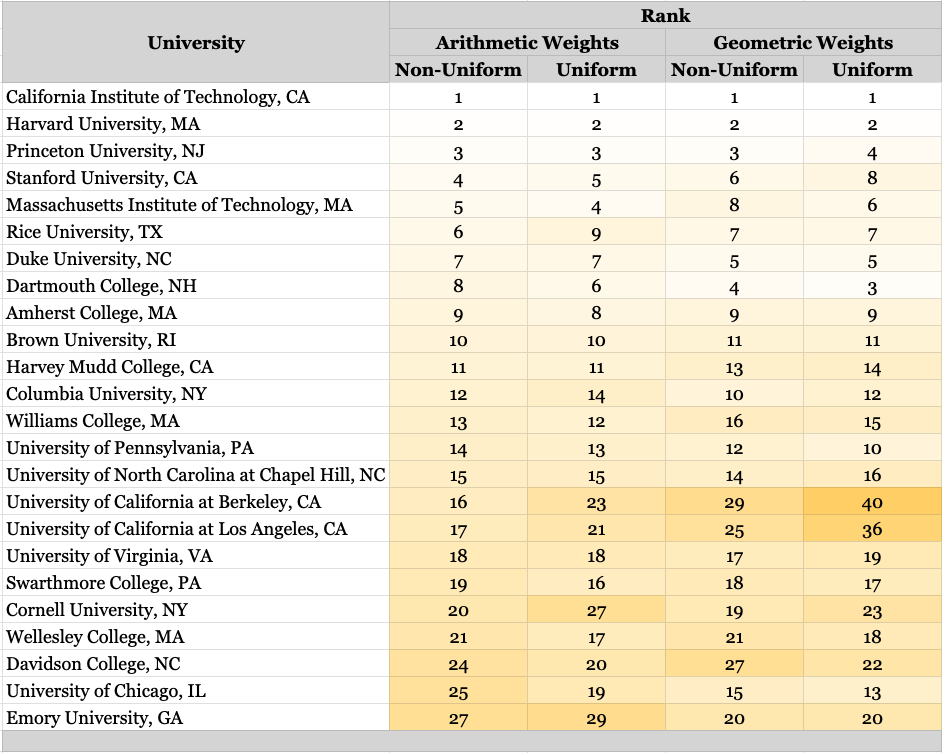}
  \caption{Rankings for the universities in the College dataset with
    respect to different weighting schemes: Arithmetic vs. geometric
    mean formulas, and for each, uniform (identical) vs. non-uniform
    (different) weights. Note the presence of some significant changes
    in ranks for especially universities at higher ranks. Also here
    and in the sequel we use yellow color for highlighting ranks and
    scores.}
  \label{fig:weights}
\end{figure}

\subsection{Step 6: Weighting and Aggregation}
\label{sec:aggr}

As mentioned earlier, each university has the same $m$ attributes but
with potentially different values. Each attribute $a_{ij}$ is paired
with a weight $w_{j}$, which is the same across all
universities. These attribute and university pairs are manipulated in
one of the following ways to generate a score so that universities can
be ranked by their scores.

{\bf The weighted arithmetic mean formula.} The score $s_{i}$ of the
$i$th university is equal to
\begin{equation}
  s_{i} = \frac{\sum\limits_{j=1}^{m} w_{j} a_{ij}}{\sum\limits_{j=1}^m w_{j}} =
  \sum\limits_{j=1}^{m} w_{j} a_{ij}
\end{equation}
where due to normalization or by construction the sum of the weights
in the denominator is equal to 1, i.e., $\sum\limits_{j=1}^m w_{j}=1$.

{\bf The weighted geometric mean formula.} The score $s_{i}$ of the
$i$th university is equal to
\begin{equation}
  s_{i} = \exp\left(\frac{\sum\limits_{j=1}^{m} w_{j} \ln(a_{ij})}{\sum\limits_{j=1}^m
    w_{j}}\right)
  = \exp\left(\sum\limits_{j=1}^{m} w_{j} \ln(a_{ij})\right)
\end{equation}
where due to normalization or by construction the sum of the weights
in the denominator is equal to 1, i.e., $\sum\limits_{j=1}^m w_{j}=1$. Here
$\exp(\cdot)$ and $\ln(\cdot)$ are the exponential and natural
logarithm functions, respectively.

Between these, the former formula is more common although the latter
is proven to be more robust~\cite{To2014}. The former formula is a
sum-of-products formula referred to by many other names in especially
the Economics literature such as ``the composite index'' (the most
common), ``the weight-and-sum method'', ``the composite indicator'',
``the attribute-and-aggregate method'', ``the simple additive
weighting'', or ``the weighted linear combination'', e.g. see
\cite{UsSa2007}. When the sum of the weights is unity, the formula is
also equivalent to the dot product of the attribute and weight
vectors.

Now we know how to compute a score per university but how do we select
the weights? There are at least three ways of selecting
weights~\cite{GaFeGu2017,Oe2008}: a) data driven such as using
principal component analysis; b) normative such as public or expert
opinion, equal weighting, arbitrary weighting; and c) hybrid
weighting. We will use the normative weighting scheme as follows.

\begin{enumerate}
\item {\bf Non-uniform weighting.} We use the student persona in
  \S~\ref{sec:attr-select} to select 20 weights
  subjectively. Fig.~\ref{fig:attr} shows the 11 weights derived out
  of these 20. Here is our heuristic for selecting the 20 weights: By
  ranking the universities per attribute, we looked at the top 10
  universities in rank visually and how they align with the top 10
  from the four well-known rankings. We then classified the attributes
  into three strength categories: High, medium, low, where high means
  the resulting ranking aligns well (again subjectively) with the top
  10 from the four rankings whereas medium and low show less
  alignment. Finally, we decided to double the weight for each
  increment in strength. Note that we give the highest weights to the
  school selectivity (via the acceptance rate (inverted)) and the
  student selectivity (via the percent of students from top 10 in high
  school).

\item {\bf Uniform weighting.} We assign the same weight of 1 to every
  attribute. This removes any potential bias due to weight differences
  among attributes but it has its own critiques, e.g., see
  \cite{GaFeGu2017}.

\item {\bf Random weighting.} We assign a uniformly random weight to
  each attribute subject to the constraint that the sum of the weights
  is equal to 1. Random weighting and its consequences are explored in
  \S~\ref{sec:random}.
\end{enumerate}

Fig.~\ref{fig:weights} compares the ranking of the top 20 in our
dataset with respect to different weighting schemes, the arithmetic
vs. geometric formula and the non-uniform vs. the uniform
weights. Note that although the few universities are the top in each
ranking, some universities such as ``University of California (UC),
Berkeley'' have significant differences in their rankings. This
observation provides another support for our claim that weighted
rankings are not robust.

A word of caution is that even though our discussion above may imply
that our top 20 ranking is similar to the ones by the well-known
rankings in part by design, we will later prove by various random or
deterministic weight selections, including no weights, that the top 20
rankings are still similar. In other words, the design heuristic we
use above is for convenience only and it is immaterial to the
conclusions of this paper.

\subsection{Steps 7-9: Robustness, Sensitivity, and Data Analysis}

We will cover this in the experimental results.

\subsection{Step 10: Visualization of the Results}

Our top 20 ranking from the College dataset is shown in
Fig.~\ref{fig:top20ours}. Our dataset and this top 20 ranking both
include liberal arts colleges; the four rankings organizations usually
have a separate ranking for liberal arts colleges. In this figure, we
show the attribute values and the final unnormalized score. The last
row shows our weights chosen for this ranking. We hope the readers can
convince themselves that the resulting rankings, both non-uniform and
uniform cases, seem reasonable and look in close alignment with the
four well-known rankings (also see the argument in the last paragraph
in the section for Step 6). 

All in all, we hope we have provided a reasonable illustration of a
ranking process in this section. As we repeatedly mention, our
mathematical formulation and conclusions are not specific to this
dataset.

\section{The Four Well-known Rankings}
\label{sec:four-rankings}

We now briefly discuss the four well-known rankings in this
section. For each ranking, we provide a very brief history, its
ranking methodology with the list of the latest attributes and their
weights, and the top 10 universities. This section covers the rankings
of universities in the aggregate as well as in Computer Science.

\begin{figure}[ht]
  \centering
  \includegraphics[scale=0.3]{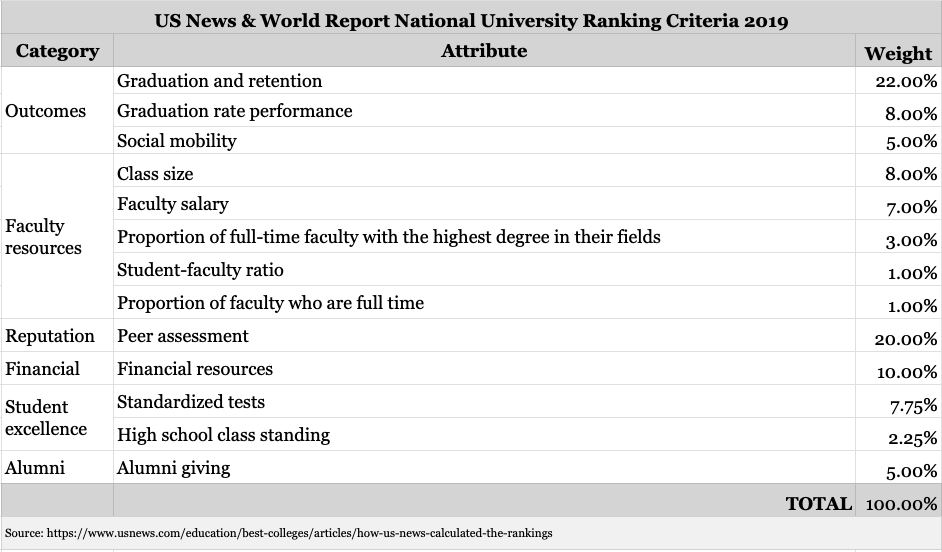}
  \caption{The attributes and weights used by US News and World Report
    in 2020 for the US national ranking.}
  \label{fig:usnews2020national}
\end{figure}

\begin{figure}[ht]
  \centering
  \includegraphics[scale=0.3]{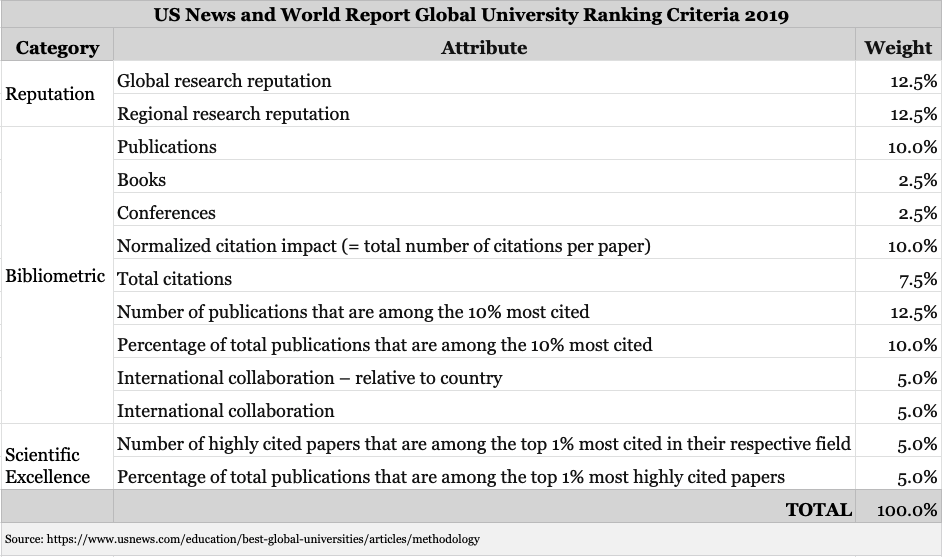}
  \caption{The attributes and weights used by US News and World Report
    in 2020 for the global ranking.}
  \label{fig:usnews2020global}
\end{figure}

\subsection{U.S. News \& World Report (USNWR) Rankings}

U.S. News \& World Report (USNWR) rankings have been active since
1985~\cite{usnews2019a}. The attributes and weights of the latest
ranking methodology are shown in Fig.~\ref{fig:usnews2020national} for
US national ranking and in Fig.~\ref{fig:usnews2020global} for global
ranking.

The national ranking has six categories of attributes, which are 13 in
total, and the weights range from 1\% to 22\%. About 20\% of the total
weight, i.e., the ``peer assessment'' attribute, is opinion based. The
global ranking has three categories of attributes, which are also 13
in total, and the weights range from 2.5\% to 12.5\%. About 25\% of
the total weight, i.e., the research reputation attributes, is opinion
based.

More details about the methodology is available at \cite{usnews2019b}
and \cite{usnews2019c} for the national and global rankings,
respectively. The latest rankings using this methodology are available
at \cite{usnews2019d} and \cite{usnews2019e} for the national and
global rankings, respectively.

\begin{figure}[ht]
  \centering
  \includegraphics[scale=0.3]{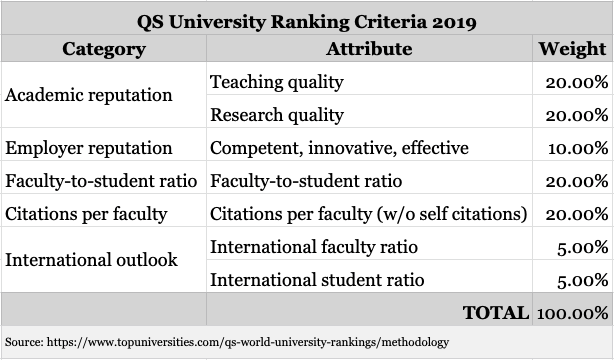}
  \caption{The attributes and weights used by Quacquarelli Symonds (QS) in
    2020.}
  \label{fig:qs2020}
\end{figure}

USNWR has rankings for objects other than universities such as
hospitals~\cite{usnews2020b} and countries~\cite{usnews2020a} using
the same weights-based ranking methodology.

\subsection{Quacquarelli Symonds (QS) Rankings}

The Quacquarelli Symonds (QS) rankings have been active since
2004~\cite{qs2019a}. Between 2004 and 2010, these rankings were done
in partnership with Times Higher Education (THE). Since 2010, QS
rankings have been produced independently. The attributes and weights
of the latest ranking methodology are shown in Fig.~\ref{fig:qs2020};
it has five categories of attributes, which are seven in total. The
weights range from 5\% to 20\%. At least 50\% of the total weight is
opinion based under the ``academic reputation'' and ``employer
reputation'' categories. More details about the methodology is
available at \cite{qs2019b}. The latest rankings using this
methodology are available at \cite{qs2019c}.

\begin{figure}[ht]
  \centering
  \includegraphics[scale=0.3]{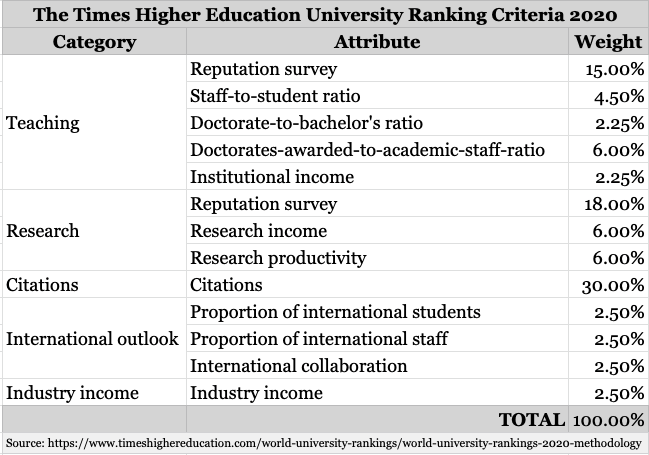}
  \caption{The attributes and weights used by Times Higher Education (THE)
    in 2019.}
  \label{fig:the2020}
\end{figure}

\subsection{Times Higher Education (THE) Rankings}

The Times Higher Education (THE) rankings have been active since
2004~\cite{the2019a}. Between 2004 and 2010, these rankings were done
in partnership with QS. Since 2010, THE rankings have been produced
independently. The attributes and weights of the latest ranking
methodology are shown in Fig.~\ref{fig:qs2020}; it has five categories
of attributes, which are 13 in total. The weights range from 2.25\% to
30\%. At least 33\% of the total weight is opinion based under the
``reputation survey'' attributes. More details about the methodology
is available at \cite{the2019b}. The latest rankings using this
methodology are available at \cite{the2019c}.

\begin{figure}[ht]
  \centering
  \includegraphics[scale=0.3]{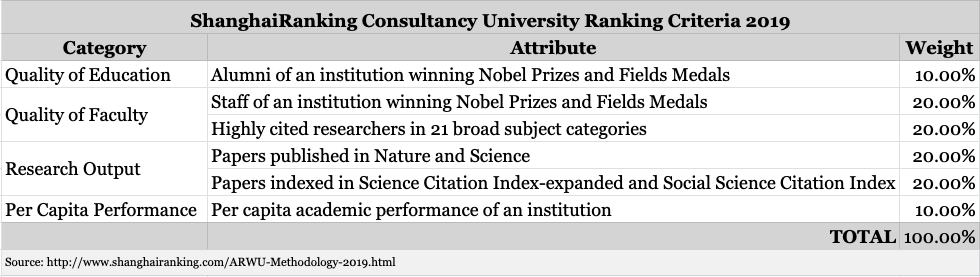}
  \caption{The attributes and weights used by ShanghaiRanking Consultancy
    in 2018.}
  \label{fig:china2020}
\end{figure}

\subsection{ShanghaiRanking Consultancy (SC) Rankings}

The ShanghaiRanking Consultancy (SC) rankings have been active since
2003~\cite{china2019a,LiCh2005}. Between 2003 and 2008, these rankings were
done by Shanghai Jiao Tong University. Since 2009, these rankings have
been produced independently. The attributes and weights of the latest
ranking methodology are shown in Fig.~\ref{fig:china2020}; it has four
categories of attributes, which are six in total. The weights range
from 10\% to 20\%. No part of the total weight is directly opinion
based. More details about the methodology is available at
\cite{china2019b}. The latest rankings using this methodology are
available at \cite{china2019c}.

\begin{figure}[ht]
  \centering
  \includegraphics[scale=0.3]{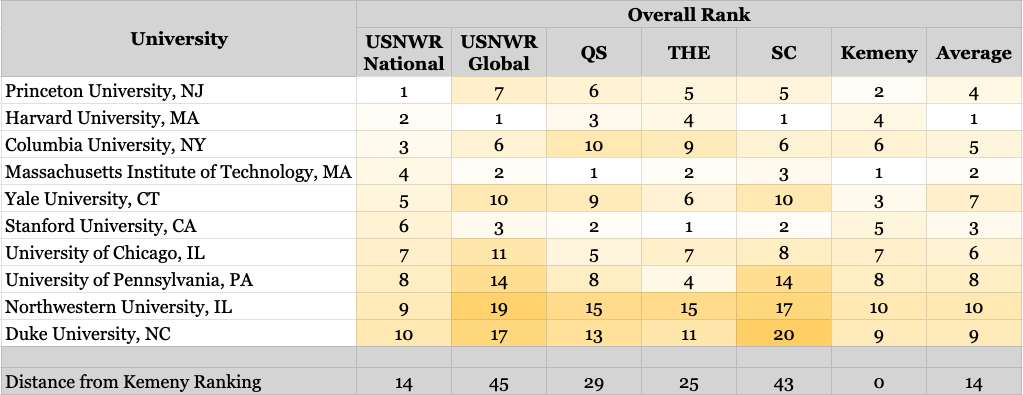}
  \caption{An illustration of the differences among the four
    well-known rankings, with the top 10 national ranking of USNWR
    taken as the reference. Also note the ranking by the Kemeny Rule
    and the average ranking computed out of the first five
    columns. The last row of numbers shows the similarity score
    between the Kemeny rule ranking and each of the other rankings, as
    measured by the Spearman's footrule distance.}
  \label{fig:top10all}
\end{figure}

\subsection{The Top 10 Overall Rankings Comparison}

To illustrate the differences between rankings,
Fig.~\ref{fig:top10all} shows the top 10 overall rankings of the US
universities. In this figure, the first column is the reference for
this table: USNWR national ranking. The next four rankings are USNWR
global ranking, and the other three well-known rankings. In the column
``Kemeny'', we present a ranking (called the Kemeny ranking) using the
Kemeny rule, which minimizes the disagreements between the first four
rankings and the final ranking~\cite{DwKuNa2001}. Finally, in the
column ``Average'', we present a ranking using the average of the
ranks over the first five columns.

Fig.~\ref{fig:top10all} already illustrates the wide variation between
these rankings: a) there is no university that has the same rank
across all these rankings; b) there is not even an agreement for the
top university; c) some highly regarded universities, e.g., UC
Berkeley, are not even in top 10 in these rankings; and d) the two
USNWR rankings of the same universities do not agree. The last row
shows the difference between each ranking and the Kemeny ranking,
where the difference is computed using Spearman's
footrule~\cite{Da2018,Sp1906}, which is nothing more than the sum of
the absolute differences between pairwise ranks. The distance shows
that the rankings closer to the Kemeny ranking in decreasing order are
the following rankings: Average, USNWR national, Times THE, QS,
Shanghai SC, and USNWR global.

It is instructive to see the top ranked university in each ranking:
Princeton University in USNWR national ranking, Harvard University in
USNWR global ranking and SC ranking, Massachusetts Institute of
Technology in QS ranking and Kemeny ranking, Stanford University in
THE ranking, Harvard University in SC ranking and the average
ranking. The top ranked university in these rankings may also change
from year to year. These disagreements even for the top ranked
university hopefully convinces the readers about the futility of
paying attention to the announcements of the top ranked university
from any rankings organization.

\begin{figure}[ht]
  \centering
  \includegraphics[scale=0.3]{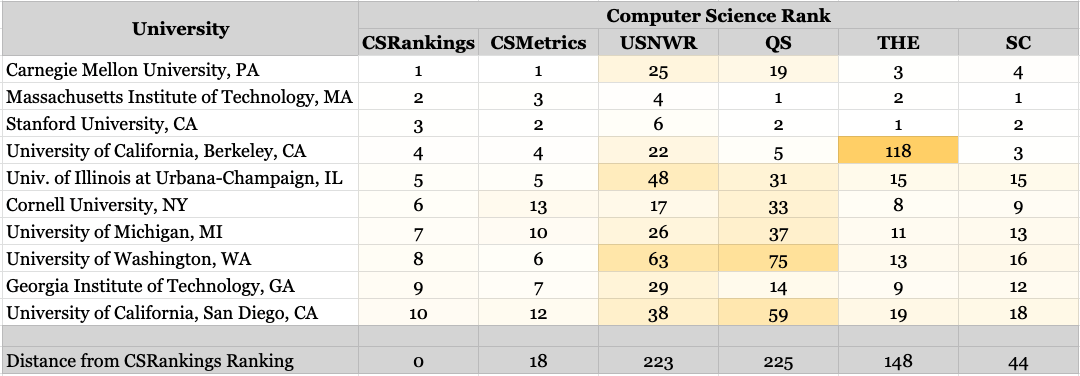}
  \caption{The top ten universities in the world in computer science
    ranking per each rankings organization. The first two rankings by
    CSRankings and CSMetrics, two computer science focused rankings
    developed and maintained by academicians. The first column is the
    reference ranking.}
  \label{fig:top10cs}
\end{figure}

\subsection{The Top 10 Computer Science Rankings Comparison}

To illustrate the differences between rankings hopefully better,
Fig.~\ref{fig:top10cs} gives the top 10 university rankings for
computer science. Three words of caution here are a) that these
rankings organizations use different titles for their computer science
rankings (USNWR: ``Computer and Information Sciences'', QS: ``Computer
Science and Information Systems'', THE: ``Computer Science'', SC:
``Computer Science and Engineering''); b) that due to these different
titles these rankings possibly cover more than computer science; and
c) that it is not clear what changes these organizations made in their
generic ranking methodologies for computer science and, for that
matter, for other subjects or areas.

The first two columns in Fig.~\ref{fig:top10cs} are by
CSRankings~\cite{cs2019b} and CSMetrics~\cite{cs2019a}, respectively,
two computer science focused rankings developed and maintained by
academicians in computer science~\cite{goto2019}. These two rankings
also align well with our own experiences as computer scientists. For
the sake of simplicity, we will refer to them as the ``academic''
rankings.

These academic rankings are mainly based on citations in almost all
the venues that matter to computer science. The central premise of
these rankings is ``to improve rankings by utilizing more objective
data and meaningful metrics''~\cite{goto2019}. These rankings intent
to follow the best practices set by the Computing Research Association
(CRA): “CRA believes that evaluation methodologies must be data-driven
and meet at least the following criteria: a) Good data: have been
cleaned and curated; b) Open: data is available, regarding attributes
measured, at least for verification; c) Transparent: process and
methodologies are entirely transparent; and d) Objective: based on
measurable attributes.” These best practices are the reason for these
sites declaring themselves as GOTO-ranking compliant, where GOTO
stands for these four criteria. For computer science rankings, a call
to ignore the computer science ranking by USNWR was made by the CRA
due to multiple problems found with the ranking~\cite{cra2019}.

Note in Fig.~\ref{fig:top10cs} the significant differences among the
rankings. As mentioned above, we agree with the academic
rankings. However, it is difficult to agree with the ranks assigned to
some universities in the other rankings. For example, it is difficult
for us to agree with Carnegie Mellon University having rank 25 in
USNWR and University of California, Berkeley having rank 118 in
THE. Any educated computer scientist would agree that these two
universities are definitely among the best in computer science. These
two examples alone show the unreliability of the ``non-academic''
rankings at least for computer science.

\section{Related Work}
\label{sec:related}

There is a huge literature on rankings, especially in the Economics
literature for rankings of countries for various well-being
measures. As a result, we cannot be exhaustive here; we will instead
refer to a fairly comprehensive set of key papers that are mainly
overview or survey papers or papers that are directly relevant to our
work.

Recall the following acronyms that we defined above for the four
well-known rankings organizations: The US News \& World Report
(USNWR), Quacquarelli Symonds (QS), Times Higher Education (THE), and
ShanghaiRanking Consultancy (SC).

\cite{KiEb2016} gives a history of rankings. \cite{Oe2008} is the de
facto bible of all things related to composite indices. Although the
focus is on well-being indices for populations and countries, the
techniques are readily applicable to university rankings, as we also
briefly demonstrated in this paper. \cite{UsSa2007} surveys 18
rankings worldwide. It acknowledges that there is no single definition
of quality, as seen by the different sets of attributes and weights
used across these rankings. It recommends quality assurance to enable
better data collection and reporting for improved inter-institutional
comparisons.

\cite{Ih2007} provides an insider view of USNWR
rankings. \cite{LiCh2005} is a related paper but on SC rankings.
\cite{SaHo2008}, though focusing on SC and THE only, follows a
general framework that can be used to to compare any two university
rankings. It finds out that SC is only good for identifying top
performers and only in research performance, that THE is undeniably
biased towards British institutions and inconsistent in the relation
between subjective and objective attributes. \cite{BiBoVi2010}
proposes a critical analysis of SC, identifies many problems with SC,
and concludes that SC does not qualify as a useful guide to neither
academic institutions nor parents and students.

\cite{Eh2005} presents and criticizes the arbitrariness in university
rankings. \cite{vRa2005} focuses on the technical and methodological
problems behind the university rankings. By revealing almost zero
correlation between the expert opinions and bibliometric outcomes, this
paper casts a strong doubt on the reliability of expert-based
attributes and rankings. This paper also argues that a league of
outstanding universities in the world may not exceed 200 members,
i.e., any ranks beyond 200 are potentially arbitrary. \cite{BoDu2015}
presents a good discussion of the technical pitfalls of university
ranking methodologies.

\cite{Cl2002} provides guidelines on how to choose
attributes. \cite{GaFeGu2017} reviews the most commonly used methods
for weighting and aggregating, including their benefits and
drawbacks. It proposes a process-oriented approach for choosing
appropriate weighting and aggregation methods depending on expected
research outcomes. \cite{DeLu2013} categorizes the weighting
approaches into data-driven, normative, and hybrid and then discusses
a total of eight weighting approaches along these categories. It
compares their advantages and drawbacks.

\cite{At2015} uses Kemeny rule based ranking to avoid the weight
imprecision problem.  \cite{Ro2013} provides a comparative study of
how to provide rankings without explicit and subjective weights. These
rankings work in a way similar to Kemeny rule based ranking.

\cite{GrIsTa2019} provides a synopsis of the choices available for
constructing composite indices in light of recent
advances. \cite{Pa2009} provides a literature review and history on
research rankings and proposes the use of bibliometrics to strengthen
new university research rankings.

\cite{Ku2014} provides an example of how to game the rankings system,
with multiple quotes from USNWR and some university presidents on how
the system works. \cite{JaSaJa2019} presents a way to optimize the
attribute values to maximize a given university's rank in a published
ranking.

\cite{DeGrLi2014,DeGrLi2019} construct a model to clarify the
incentives of the ranker, e.g., USNWR, and the student. They find the
prestige effect pushing a ranker into a ranking away from
student-optimal, i.e., not to the advantage of the student. They
discuss why a ranker chooses the attribute weights in a certain way
and why they change them over time. They also present a
student-optimal ranking methodology. \cite{LuSm2015} exposes the games
business school play whether or not to reveal their rankings.

\cite{LuSm2013} provides the casual impact of rankings on application
decisions, i.e., how a rank boost or decline of a university affect
the number of applications the university gets in the following year.

University rankings are an instance of multi-objective
optimization. \cite{ChWaCh2017} provides a survey of such systems
spanning many different domains, including university
rankings. \cite{GuAsMa2019} presents applications to ranking in
databases.

In summary, rankings like many things in life have their own pros and
cons~\cite{Gl2011}. The pros are that they in part rely on publicly
available information~\cite{AgDeHo2008}; that they bring attention to
measuring performance~\cite{AgDeHo2008}; that they have provided a
wake-up call, e.g., in Europe~\cite{AgDeHo2008}, for paying attention
to the quality of universities and providing enough funding for them
due to the strong correlation between funding and high rank; that they
provide some guidance to students and parents in making university
choices; that they use easily understandable attributes and weights
and a simple score-based ranking.

The cons are unfortunately more than the pros. The cons are that data
sources can be subjective~\cite{vRa2005}, can be and has been
gamed~\cite{Ri2019}, can be incomplete; that attributes and weights
sometimes seem arbitrary~\cite{AgDeHo2008,Pe2011}; that weights accord
too little importance to social sciences and
humanities~\cite{AgDeHo2008}; that many operations on attributes and
weights affect the final rankings~\cite{GrIsTa2019}; that many
attributes can be highly correlated~\cite{Pe2011}; that there is no
clear definition of quality~\cite{UsSa2007}; that rankings encourage
rivalry among universities and strengthen the idea of the academic
elite~\cite{vRa2005}; that rankings lead to a ``the rich getting
richer'' phenomenon due to highly ranked universities getting more
funding, higher salaries for their admin staff, more demand from
students, and more favorable view of quality in expert
opinions~\cite{vRa2005}; that assigning credit such as where an award
was given vs. where the work was done is unclear~\cite{vRa2005}; that
expert opinions are shown to be statistically unreliable and yet some
well-known ranking organizations still use them~\cite{vRa2005}; that
even objective bibliometric analysis has its own
issues~\cite{vRa2005}; that some rankings such as THE have undeniable
bias towards British universities~\cite{SaHo2008}; and that the
current rankings cannot be trusted~\cite{BiBoVi2010,Eh2005,goto2019}.

The current well-known rankings have created their own industry and
they have strong financial and other incentives to continue their way
of presenting their own rankings~\cite{DeGrLi2019}. There are
initiatives to close many drawbacks of the current rankings such as
the ``Common Data Set Initiative'' to provide publicly available data
directly from universities~\cite{cds2019}, Computer Science rankings
created by people who know Computer Science as in academicians from
Computer Science~\cite{goto2019}, a set of principles and requirements
(called the Berlin Principles) that a ranking needs to satisfy to
continuously improve~\cite{Berlin2019}, an international institution to
improve the rankings methodology~\cite{Ireg2019}, extra validation
steps and prompt action by the current rankings organization against
gaming~\cite{Ih2007,Ri2019}. In short, there is hope but it will take
time to reach a state where many of the pros have been eliminated.

\section{Mathematical Formulation}
\label{sec:math}

We have $n$ objects to rank. Objects are things like universities,
schools, and hospitals. A ranking is presented to people to select one
of. Each object $i$ has $m$ numerical attributes from $a_{i1}$ to
$a_{im}$, index with $j$. We can use the matrix $A$ to represent the
objects and their attributes, one row for each thing and one column
for each attribute.
\begin{equation}
  \label{eq:A}
A=\left[
\begin{matrix}
  a_{11} & a_{12} & \cdots & a_{1m} \\
  a_{11} & a_{12} & \cdots & a_{1m} \\
  \vdots & \vdots & \vdots & \vdots \\
  a_{n1} & a_{n2} & \cdots & a_{nm} \\  
\end{matrix}
\right]
\end{equation}

We have $m$ unknown numerical coefficients from $w_{1}$ to $w_{m}$ in
the vector $w$.
\begin{equation}
  \label{eq:w}
w=\left[
  \begin{matrix}
    w_{1} & w_{2} & \cdots & w_{m}
  \end{matrix}
  \right],
\end{equation}
where each $w_{j}$ represents a weight for the attribute $a_{ij}$ for
some $i$. Each weight is non-negative. The same weights are used for
every row of attributes. The sum of the weights is set to 1 due to
normalization.

For each object $i$, we compute a score $s_{i}$ as
\begin{equation}
  \label{eq:score}
  s_{i} = \sum\limits_{j=1}^{m} w_{j} a_{ij}
\end{equation}
or in the matrix form
\begin{equation}
  \label{eq:matrix}
  s = Aw
\end{equation}
where $A$ is the matrix of the known attributes and $w$ is the set of
unknown weights, both as defined above. Recall from \S~\ref{sec:aggr},
the score formula above is the arithmetic mean formula, and the scores
can also be computed using the geometric mean formula. Note that $Aw$
is a matrix-vector multiplication. Also note that setting each weight
in $w$ to 1 is the uniform weight case.

Fig.~\ref{fig:top20ours} illustrates the attributes, weights, scores,
and ranks for the top 20 universities from the College dataset. The
attributes across all 20 rows and 11 columns represent the matrix $A$
for the top 20 universities only; the actual matrix $A$ has over 600
rows. The last row in this table represents the vector $w$ of 11
weights we assigned for this ranking. The last two columns represent
the vector $s$ of scores and the ranks for each university.

{\bf Ordering.} The top $n$ ranking, like the top 20 in
Fig.~\ref{fig:top20ours}, is an ordering of these universities in
decreasing score, i.e.,
\begin{equation}
  \label{eq:order}
  s_{1} > s_{2} > \cdots > s_{n},
\end{equation}
which we will refer as the score ordering constraint. Another form of
these inequalities is
\begin{equation}
  s_{i} \geq s_{i + 1} + \epsilon,\;\;\forall i\in [1,n-1],
\end{equation}
where $\epsilon$ is a small constant. This form will be useful in
linear programming formulations.

Later in \S~\ref{sec:kemeny} we will introduce another order of
universities without using their scores.

{\bf Domination.} Given two attribute vectors $a_{x}$ and $a_{y}$ of
length $m$, we say $a_{x}$ strictly dominates $a_{y}$ if and only if
for all $j$, $a_{x}[j] \geq a_{y}[j]$; we say $a_{x}$ partially
dominates $a_{y}$ if and only if $a_{y}$ does not strictly dominate
$a_{x}$. Note that if $a_{x}$ does not strictly dominate $a_{y}$, then
it is necessarily true that $a_{x}$ and $a_{y}$ partially dominate
each other for different sets of their attributes.

The domination idea is due to the impact on the score ordering. Given
two attribute vectors $a_{x}$ and $a_{y}$, it is easy to see that if
$a_{x}$ partially dominates $a_{y}$, we can always find a set of
weights to make $s_{x} > s_{y}$, calculated as in
Eq.~\ref{eq:score}. 

{\bf (Integer) Linear Programming (LP) Formulation.} In the next
section we will define a set of problems. For each problem we will
formulate a linear program or an integer linear program and solve it
using one of the existing open source LP packages. For our
experiments, the LP package we used was
lp\_solve~\cite{lpsolve2019}. Details are in the following problems
and solutions section. For the constraints of these programs, strict
domination is used extensively. Note that although ILP is NP-hard, it
takes a few seconds on a personal laptop to generate our ILP programs
using the Python programming language and run them using our LP
package for over 600 universities. As a result, we do not see any
reason (potentially other than intellectual curiosity) to develop
specialized algorithms for the problems we study in this paper.

\section{Explorations of Different Rankings}
\label{sec:explore}

We will now show that there are multiple valid choices in assigning
ranks to universities. For each choice, we will pose a problem and
then provide a solution to it. Below there will be a section dedicated
to each problem.

In Problem 1, we explore the existence of different rankings using
Monte Carlo simulation. We will do so in two ways: a) how many
universities can be moved to rank 1, b) whether or not we can find
weights to keep a given top k ranking of universities. The search
space here is the space of weight vectors, called the weight space.

Our solution to Problem 1 may have two issues due to the use of
simulation: The weight space may not be searched exhaustively and the
search may not be efficient. Using linear programming, Problem 2
ensures that the weight space exploration is both efficient and
optimal.

In the first two problems the existence of different rankings reduces
to the existence of different weights under ordering constraints. As
long as such weights exist, we are concerned about how they may appeal
to a human judge. In Problem 3, we rectify this situation in that we
derive weights that would appeal to a human judge as reasonable or
realistic as if assigned by a human.

These three problems show that there are many universities that can
attain the top rank but not every university can achieve it. Then a
natural next question to ask is how to find the best rank that each
university can attain. Problem 4 is about solving this problem.

The first four problems always involve attribute weights. In Problem
5, we explore the problem of finding rankings without using weights at
all. The solution involves a technique of aggregating rankings
per-attribute, called the Kemeny rule.

In Problem 6, we explore the problem of how much improvement in the
ranking of a given university is possible by improving attribute
values in a weight-based scoring methodology. This problem should
provide some guidance to universities in terms of what to focus on
first to improve their ranks.

\begin{figure}[ht]
  \centering
  \includegraphics[scale=0.3]{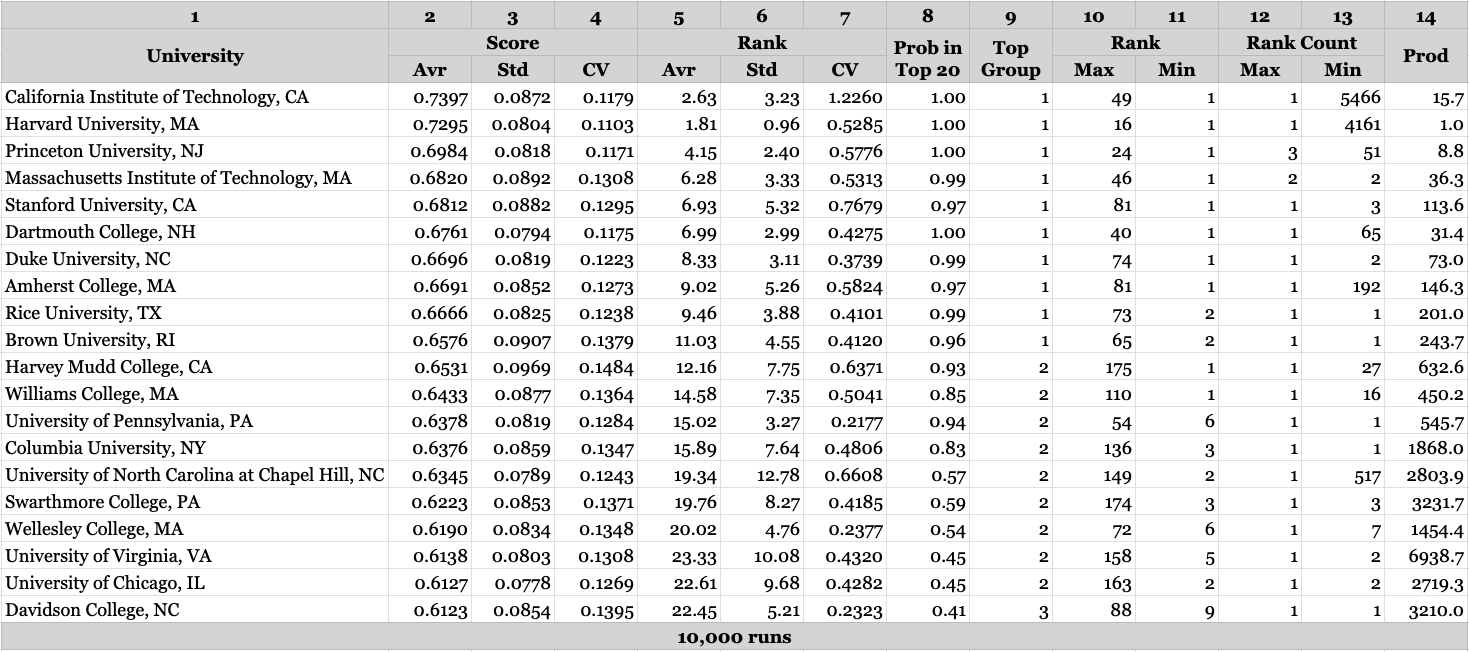}
  \caption{Ranking of the top 20 universities in decreasing average
    scores order using the arithmetic mean formula and using uniformly
    random weights in all 10,000 runs. ``Avr'', ``Std'', and ``CV'' in
    Columns 2-7 stand for the average, the standard deviation, and the
    coefficient of variation, respectively. ``Prob in Top 20'' in
    Column 8 is the probability of falling in the top 20 universities
    when they are ranked in decreasing score order. Each column is
    explained in \S~\ref{sec:random}.}
  \label{fig:random-arith}
\end{figure}

\begin{figure}[ht]
  \centering
  \includegraphics[scale=0.3]{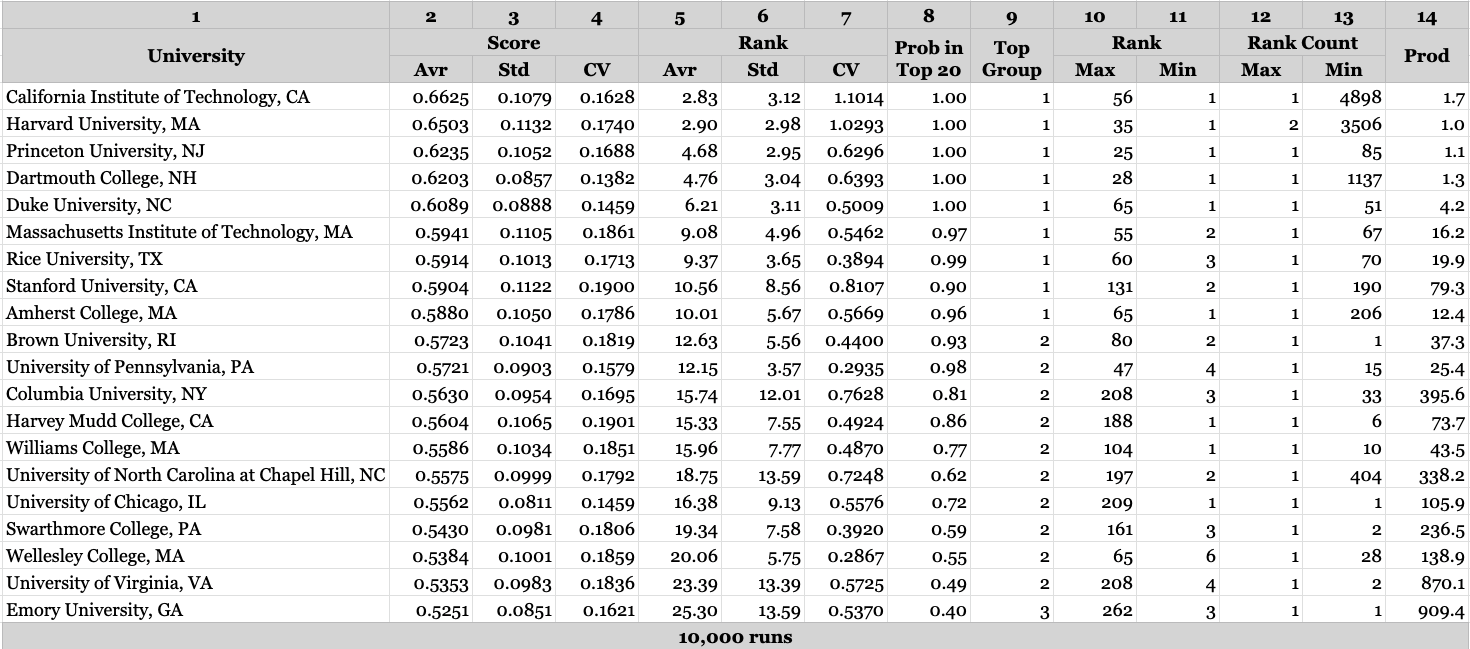}
  \caption{Ranking of the top 20 universities in decreasing average
    scores order using the geometric mean formula and using uniformly
    random weights in all 10,000 runs. ``Avr'', ``Std'', and ``CV'' in
    Columns 2-7 stand for the average, the standard deviation, and the
    coefficient of variation, respectively. ``Prob in Top 20'' in
    Column 8 is the probability of falling in the top 20 universities
    when they are ranked in decreasing score order. Each column is
    explained in \S~\ref{sec:random}.}
  \label{fig:random-geo}
\end{figure}

\subsection{Problem 1: Rankings with Uniformly Random Weights}
\label{sec:random}

Assigning weights subjectively has its own issues so what if we do not
assign weights manually at all? In this section, we explore the weight
space automatically to discover different rankings and find out what
extremes are possible.

Since we use randomization, we need repetition to get meaningful
outcomes on the average. Let $N$ be the number of runs. In our
experiments, we set $N$ to 10,000. Each run derives $m$ weights with
three constraints: a) each weight is independently and identically
drawn (iid); b) each weight is uniformly random; c) the sum of the
weights is equal to 1. Among these constraints, care is needed for
constraint c, for which we adapted an algorithm suggested in
\cite{An2015}.

Over these $N$ runs, we collect the following results for each
university:
\begin{enumerate}
  \setcounter{enumi}{1}
\item {\em Average (Avr) of scores},
\item {\em Standard deviation (Std) of scores},
\item {\em Coefficient of variation (CV) of scores}, computed as the
  ratio of the standard deviation to the average,
\item {\em Average of ranks},
\item {\em Standard deviation of ranks},
\item {\em Coefficient of variation of ranks},
\item {\em Probability of falling into Top 20},
\item {\em Top group id}, explained in \S~\ref{sec:reco},
\item {\em Maximum rank} attained,
\item {\em Minimum rank} attained,
\item {\em Maximum count}, the number of times the maximum rank has
  been attained,
\item {\em Minimum count}, the number of times the minimum rank has
  been attained,
\item {\em Product}, explained in \S~\ref{sec:reco},
\end{enumerate}
where the line numbers indicate the column numbers in
Fig.~\ref{fig:random-arith} and Fig.~\ref{fig:random-geo}. We
generated these results for both the arithmetic and geometric means.

The top 20 rankings for both the arithmetic and geometric mean
formulas are shown in Fig.~\ref{fig:random-arith} and
Fig.~\ref{fig:random-geo}, respectively. In these figures, the columns
right after the university names are the ranks by decreasing average
score. The rest of the columns map to the list of the results above in
order.

A couple of observations are in order.
\begin{itemize}
  \item The average score and average rank rankings are
    probabilistically identical to those by the uniform weight. This
    is because the constraint c above ensures that the expected value
    of each weight is equal to the uniform weight (which is easy to
    prove using the linearity of expectation together with the
    constraints a-c).
  \item The maximum rank values show that there were runs in which
    every university was not in the top 20 but as small maximum rank count
    values together with small average rank values show that these max
    rank values were an extreme minority. More specifically, in the
    arithmetic case, Harvard University was the best as it did not
    drop below rank 16 in all $N$ runs while in the geometric case,
    Princeton University was the best as it did not drop below rank 25
    in all $N$ runs. At the same time, with respect to the average
    score ranking, neither of these universities was at rank
    1. Moreover, in the arithmetic case, Harvard University had the
    smallest average rank while in the geometric case Princeton
    University did not have the smallest average rank.
  \item The minimum rank values show that about half of the universities
    never reached the top rank over all $N$ runs. From the opposite
    angle, this also means that about half of the universities were at
    the top position in some runs.
  \item Column 13 tells us how many times a university was at its
    minimum rank. It seems some universities do reach the top position
    but it is rare. On the other hand, for the top two universities,
    the top position is very frequent. More specifically, in the
    arithmetic case, California Institute of Technology is at rank 1
    in about 55\% of the runs whereas Harvard University is at rank 1
    for about 42\%. In the geometric case, again these two
    universities has the highest chance of hitting the top position at
    about 49\% and 35\%, respectively.
\end{itemize}

These observations conclusively show that a single ranking with
subjective or random weights is insufficient to assert that a
particular university is the top university or at a certain
rank. Moreover, it is unclear which metric is the definitive one to
rank these universities; we could as well rank these universities per
average score, average rank, maximum rank reached, minimum rank
reached, or probability of hitting a certain rank, each yielding a
different ranking. We will return to these possibilities in
\S~\ref{sec:reco}.


\subsection{Problem 2: The Feasibility of Different Rankings}
\label{sec:feasible}

A large number of runs in Problem 1 may explore the weight space quite
exhaustively but the exploration using simulation still cannot be
guaranteed to be fully exhaustive. Moreover, the search itself may not
be efficient due to direct dependence on the number of runs. In this
section, we use LP to guarantee optimality and efficiency.

We have two cases: The special case is to enforce the top 1 rank for a
single university whereas the general case is to enforce the top k,
from 1 to k, for a given top k universities in order. The input is our
ranking of all the universities in our dataset using geometric mean
formula with uniform weights.

{\bf Enforcing for top 1 rank.} This case asks whether or not a set
$w$ of weights exists to ensure a given university strictly dominates
every other university. This is done by moving the given university to
rank 1 and checking if $s_{1}$ is greater than every other score. In
this problem, we are not interested in finding $w$, although LP will
return it, rather we are interested in its existence.

Let us see how we can transform this special case into a linear
program. The special case requires that $s_{1} > s_{i}$ for any $i\geq
2$, or equivalently, $s_{1}-s_{i} > 0$ for any $i\geq 2$. Since both
$w$ and $s$ are unknown, a linear program cannot be created. However,
if we subtract each row of attributes (component-wise) from the first
row, we convert $s=Aw$ into $Dw > 0$ where
\begin{equation}
  \label{eq:d}
  row_{i-1}(D)=row_{1}(A)-row_{i}(A)
\end{equation}
for $i\geq 2$, where $row_i(\cdot)$ represents the $i$th row of its
argument matrix. This converts the $n$ rows in $A$ into $n-1$ rows in
$D$. The resulting program in summary is
\begin{equation}
  w \geq 0, \sum\limits_{j=1}^{m} w_{j} = 1, Dw > 0, 
\end{equation}
where the equality constraint on the sum of the weights is enforced to
avoid the trivial solution $w=0$. This linear program can be rewritten
more explicitly as
\begin{equation}
  \label{eq:p2linear1}
  \begin{array}{ll}
  \mbox{minimize } 1 & \\
  \mbox{subject to } & \\
  (a) \; w_{j} \geq 0 \;\;(\forall j\in [1,m]), & \\
  (b) \; \sum\limits_{j=1}^{m} w_{j} = 1, & \\
  (c) \; (a_{1j}-a_{ij})w_{j} \geq\epsilon \;\; (\forall i\in [2,n]
  \mbox{ and } \forall j\in [1,m]), & \\
  \end{array}
\end{equation}
where the lower bound $\epsilon$ is set to zero or a small nonzero
constant, 0.05\% in our experiments. The zero lower bound case allows
ties in ranking whereas the nonzero lower bound case enforces strict
domination. Note that this linear program has a constant as an
objective function, which indicates that a feasibility check rather
than an optimization check is to be performed by the LP package we are
using.

Now if this linear program is feasible, then this implies there exists
a set of weights $w$ that satisfy all these constraints, or
equivalently, our special case has a solution.

The results of this experiment for the special case are as follows. We
generated and solved the linear program for each of the 609
universities in our dataset. How many universities could be moved to
the top rank? For the zero and nonzero lower bound cases, the numbers
are 45 and 28, respectively.

It is probably expected that multiple of the top ranked universities
could be moved to the top rank. For example, for the zero and nonzero
lower bound cases, any of the top 13 and 4, respectively, could
achieve the top rank. What was surprising to find out that some
universities at high ranks could also be moved to the top rank; for
the zero and nonzero lower bound cases, the highest ranks were 553 and
536, respectively.

One note on the difference in the findings between Monte Carlo
simulation vs. LP. Our Monte Carlo simulation was able to find about
12 universities that could be moved to the top rank, whereas LP was
able to find more, as given above, in a fraction of the time;
moreover, the 12 universities found by Monte Carlo simulation were
subsumed by the ones found by LP. Although this is expected due to the
LP's optimality guarantee, it is worth mentioning to emphasize the
importance of running an exhaustive but efficient search like LP.

{\bf Enforcing top k ranks.} For the general case, we need to enforce
more constraints. We require strict domination in succession, i.e.,
$s_{i}-s_{i+1} > 0$ for $i=1$ to $k-1$, to enforce the rank order for
the top $k$ universities. The resulting linear program is
\begin{equation}
  \label{eq:p2lineark1}
  \begin{array}{ll}
  \mbox{minimize } 1 & \\
  \mbox{subject to } & \\
  (a) \; w_{j} \geq 0 \;\;(\forall j\in [1,m]), & \\
  (b) \; \sum\limits_{j=1}^{m} w_{j} = 1, & \\
  (c) \; (a_{ij}-a_{i+1,j})w_{j} \geq\epsilon \;\; (\forall i\in
       [1,k-1] \mbox{ and } \forall j\in [1,m]), & \\
  (d) \; (a_{kj}-a_{ij})w_{j} \geq\epsilon \;\; (\forall i\in [k+1,n]
       \mbox{ and } \forall j\in [1,m]), & \\
  \end{array}
\end{equation}
where $k$ is a given constant less than $n$.

Taking the ranking in Fig.~\ref{fig:top20ours} as input, we wanted to
find out the maximum k such that we can find a set of weights to
enforce the top k ranking. For the zero bound case, we could find such
weights for each k from 1 to 20. For the nonzero bound case, we could
find such weights for each k from 1 to 18.

We could increase these k even further by figuring out which
universities need to move up or down in the ranking. To find them out,
we used a trick suggested in \cite{lpinfeasible2019}. This trick
involves adding a slack variable to each inequality in
Eq.~\ref{eq:p2lineark1} and also adding their sum with large numeric
coefficients in the objective function. The goal becomes discovering
the minimum number of nonzero slack variables. For each nonzero
slack variable, the implication is that the ordering needs to be
reversed. The LP formulation is below for reference:
\begin{equation}
  \label{eq:p2lineark2}
  \begin{array}{ll}
  \mbox{minimize } \sum\limits_{i=1}^{k-1} M d & \\
  \mbox{subject to } & \\
  (a) \; w_{j} \geq 0 \;\;(\forall j\in [1,m]), & \\
  (b) \; \sum\limits_{j=1}^{m} w_{j} = 1, & \\
  (c) \; (a_{ij}-a_{i+1,j})w_{j} + d_{i} \geq\epsilon \;\; (\forall
  i\in [1,k-1] \mbox{ and } \forall j\in [1,m]), & \\
  (d) \; (a_{kj}-a_{ij})w_{j} + d_{i} \geq\epsilon \;\; (\forall i\in
  [k+1,n] \mbox{ and } \forall j\in [1,m]), & \\
  \end{array}
\end{equation}
where $M$ is a large integer constant like 1,000 and $d$ are the slack
variables. Since the point about top k for a reasonable k is already
made, we will not report the results of these experiments.

The significance of these experiments is that a desired ranking of top
k for many values of k can be enforced with a suitable selection of
attributes and weights. This is another evidence for our central
thesis that university rankings can be unreliable.

\subsection{Problem 3: Appealing Weights}
\label{sec:appealing}

When we explored in the problems above the existence of weights to
enforce a desired ranking for top k, we did not pay attention to how
these weights look to a human judge. It is possible that a human judge
may think she or he would never assign such odd looking weights, e.g.,
very uneven distribution of weight values or weights that are too
large or too small. Although such an objection may not be fair in all
cases, it is a good idea to propose a new way of deriving weights that
are expected to be far more appealing to human judges. In this
section, we will explore this possibility.

Our starting point is the claim that uniform weighting removes most or
all of the subjectivity with weight selection. This claim has its own
issues as discussed in the literature but we feel it is a reasonable
claim to take advantage of. We can use this claim in two ways: a) create
rankings using uniform weights, b) approximate uniform weights. The
latter is done with the hope that it can generate rankings with larger
k.

The results with uniform weights are given in
Fig.~\ref{fig:weights}. In this section we focus on approximating
uniform weights. Our approximation works by minimizing the difference
$d$ between the maximum derived weight and the minimum derived weight
so that the weights are closer to uniform as the difference gets
closer to zero as shown below:
\begin{equation}
  \label{eq:diff}
  d = \max_{j=1}^{m} w_j - \min_{j=1}^{m} w_j,
\end{equation}
where we will refer to the numerator and denominator in this equation
as $max_w$ and $min_w$, respectively, so that we can use them as
parameters in our linear program.

Using the most basic properties of the maximum and minimum functions
as in
\begin{equation}
  max_w \geq w_j \mbox{ and } w_j\geq min_w
\end{equation}
for each $j$ from 1 to $m$, our linear program is
\begin{equation}
  \label{eq:p3linear}
  \begin{array}{ll}
  \mbox{minimize } d = max_w - min_w & \\
  \mbox{subject to } & \\
  (a) \; w_{j}\leq max_w & (\forall j\in [1, m]), \\
  (b) \; w_{j}\geq min_w & (\forall j\in [1, m]), \\
  (c) \; min_w \geq 0, \\
  (d) \; \sum\limits_{j=1}^{m} w_{j} = 1, & \\
  (e) \; Dw \geq \epsilon, & \\
  \end{array}
\end{equation}
where $Dw$ is to be defined below for each case.

\begin{figure}[ht]
  \centering
  \includegraphics[scale=0.3]{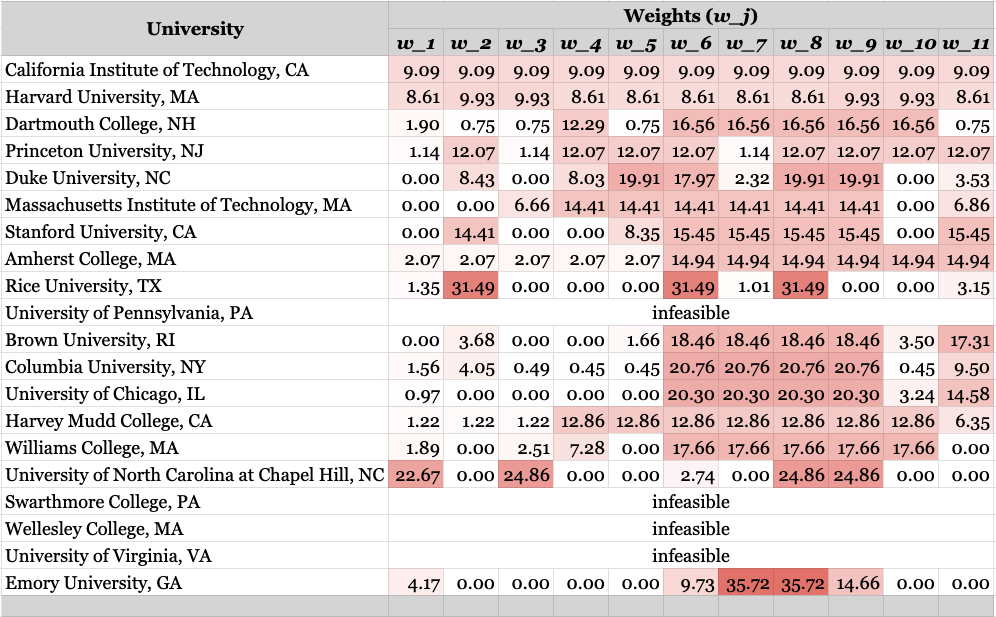}
  \caption{Weights (in percentages) to guarantee the top 1 for each
    university in our top 20 ranking. Here the label {\em w\_i} refers
    to $w_i$. The rows marked ``infeasible'' mean no weights could be
    found by LP. Also here and in the sequel we use red color for
    highlighting weights.}
  \label{fig:top20top1}
\end{figure}

{\bf Enforcing for top 1 rank.} The results of this experiment for
feasibility is the same as in the special case of Problem 1 (with $D$
defined as in Eq.~\ref{eq:d}), even though we changed the linear
program slightly. For the weights derived by the linear program, refer
to Fig.~\ref{fig:top20top1}.

In this figure, each row gives the set of weights that will guarantee
the rank 1 position for the university at the same row. We claim that
the derived weights would look reasonable to a human judge but we
encourage the reader to use their own judgment in comparison with the
weights used by the four rankings organizations presented earlier.

Another way of looking at these derived weights is to see which
attributes get higher weights. It is perhaps reasonable to argue that
such attributes provide strengths of the university that they
belong. Following this line of thinking, we may say that the top
ranked California Institute of Technology is strong across all its
attributes whereas the lowest ranked Emory University in our original
top 20 ranking has the ``Percent of faculty with PhD degrees'' and
``Faculty to student ratio'' as its strongest attributes. For a
prospective student, this line of thinking can provide two viewpoints:
a) the best university is the one that has most of its weights closer
to uniform, or b) the best university is the one that has its highest
weights for the attributes that the student is interested in. In our
opinion both viewpoints seem valid.

Note that four rows have ``infeasible'', meaning that no weights exist
to make the corresponding universities top ranked. These can also be
confirmed to some extent via the simulations as shown in
Fig.~\ref{fig:random-arith} and Fig.~\ref{fig:random-geo}. In those
figures, the minimum ranks these universities could reach in 10,000
simulations were never the top rank. Here we say ``to some extent''
because the coverage of LP is exhaustive while that of random
simulation is not.

\begin{figure}[ht]
  \centering
  \includegraphics[scale=0.3]{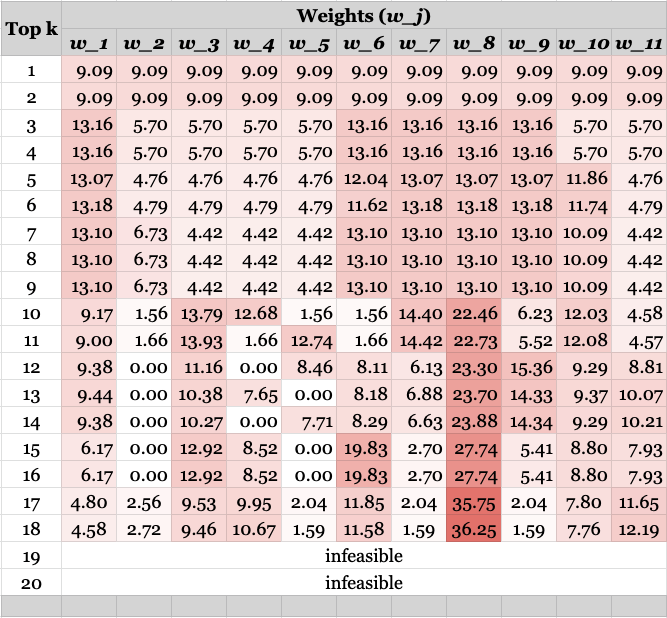}
  \caption{Weights (in percentages) to guarantee the top k for each k
    from 1 to 20. Here the label {\em w\_i} refers to $w_i$. The rows
    marked ``infeasible'' mean no weights could be found by LP.}
  \label{fig:top20topk}
\end{figure}

{\bf Enforcing top k ranks.} The results of this experiment for
feasibility is the same as in the general case of Problem 1 (with $D$
defined as in Eq.~\ref{eq:d}), even though we changed the linear
program slightly. For the weights derived by the linear program, refer
to Fig.~\ref{fig:top20topk}.

In this figure, each row gives the weights to guarantee the ranking of
top k, where k is the value in the first column of the related
row. That is, we derive the set of weights to enforce the top k
ranking as given in our top 20 rankings, for k from 1 to 20. It is not
a coincidence that top 1 weights match the first row in
Fig.~\ref{fig:top20top1}.

As in the top 1 case, we have some ``infeasible'' rows, namely, the
last two rows. This means we could find weights to enforce the ranking
up to top 18 only. For those last two rows, there exist no weights to
enforce their ranks to the 19th and 20th, respectively, unless we
reorder the universities.

\begin{figure}[ht]
  \centering
  \includegraphics[scale=0.3]{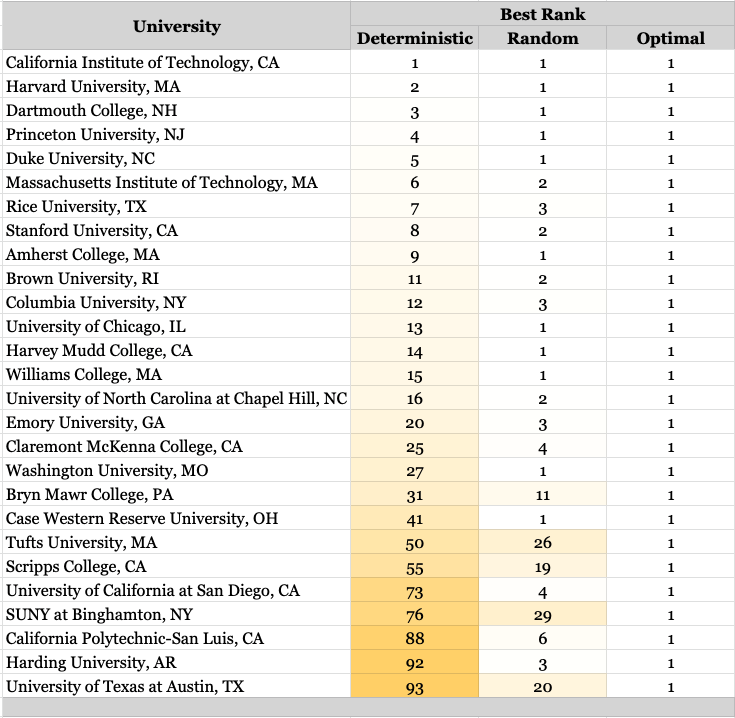}
  \caption{27 universities that reach the optimal rank of 1.}
  \label{fig:best1}
\end{figure}

\begin{figure}[ht]
  \centering
  \includegraphics[scale=0.3]{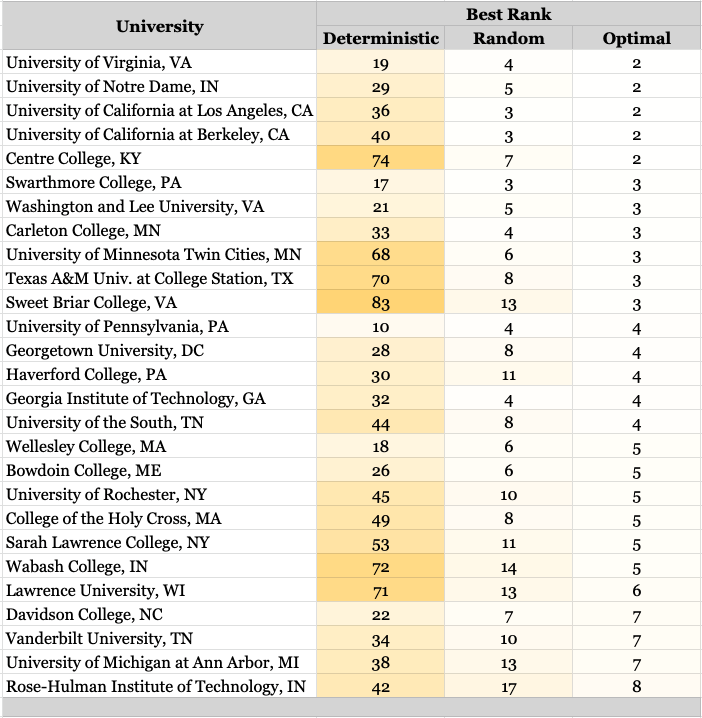}
  \caption{The next 27 universities that reach the optimal ranks up to
    8.}
  \label{fig:best2}
\end{figure}

\subsection{Problem 4: Best Possible Ranks}
\label{sec:best}

In the first three problems, we have seen that not every university
can attain the top rank or the top score. In this section, we want to
conclusively find out the top rank each university can attain.

The approach we take to compute the top rank possible for a university
works in three steps: 1) move the university to the top rank; 2)
generate the constraints to enforce that the score of the university
dominates every other university score; 3) count the number of score
constraints that are not satisfied. The last step ensures that for
every violated constraint, the enforced order was wrong and the
university in question needs to move one rank down for each
violation. In the end the count of these violations gives us the top
rank the university can attain in presence of all the other
universities in the dataset.

We again want to use LP for the approach above. The formulation is
similar to that of Eq.~\ref{eq:linearx} but we need a trick to count
the number of constraint violations. We found such a trick in
\cite{BoDu2016}, which when combined with our formulation gets the job
done. The trick involves generating two new variables $y_i$ and $d_i$
for every constraint $s_1-s_i > 0$, turning this constraint into
$s_1-s_i + d_i > 0$. If $s_1-s_i > 0$, i.e., the score constraint is
satisfied, we want the LP to set $d_i \leq 0$; on the other hand, if
$s_1-s_i \leq 0$, i.e., the score constraint is violated, we want the
LP to set $d_i > 0$. In addition, we want to count the number of
violations, i.e., the number of times $d_i > 0$. This is where $y_i$,
which can only be 0 or 1, comes into the picture. We bring $y_i$ and
$d_i$ in the form of a new constraint: $d_i - M y_i \leq 0$. For a
large constant $M$, every time $d_i > 0$, this new constraint gets
satisfied only if $y_i=1$. Every time $y_i$ is 1, this means the
university in question needs to be demoted by 1 in
rank. This also means the total number of times this demotion happens
will gives us the top rank. One caveat here is to ensure $y_i$ is zero
every time $d_i\leq 0$ but when $d_i\leq 0$, $d_i - M y_i \leq 0$ is
satisfied with $y_i$ zero or one, i.e., the latter needs to be
avoided. This is achieved with the objective function: Minimize the sum
of $y_i$, which will avoid $y_i=1$ unless it is absolutely required.

Our LP formulation implementing the approach above is
\begin{equation}
  \label{eq:linearx}
  \begin{array}{ll}
  \mbox{minimize } \sum\limits_{i=2}^{n} y_{i} & \\
  \mbox{subject to } & \\
  (a) \; w_{j} \geq 0 \;\;(\forall j\in [1,m]), & \\
  (b) \; \sum\limits_{j=1}^{m} w_{j} = 1, & \\
  (c) \; (a_{1j}-a_{ij})w_{j} + d_{i} \geq\epsilon \;\; (\forall i\in [2,n]
  \mbox{ and } \forall j\in [1,m]), & \\
  (d) \; d_{i} - M y_{i} \leq 0 \;\; (\forall i\in [2,n]), & \\
  (e) \; y_{i} \in \{0,1\} \;\; (\forall i\in [2,n]), & \\
  \end{array}
\end{equation}
where $M$ is a large enough constant, which we set to 10 in our
experiments.

Fig.~\ref{fig:best1} and Fig.~\ref{fig:best2} present the results in
two tables. The table in Fig.~\ref{fig:best1} contains the top 27
universities and the table in Fig.~\ref{fig:best2} contains the next
set of 27 universities. For each university, these tables have three
top ranks that these universities can attain: the ``Deterministic''
one coming from the geometric uniform ordering, the ``Random'' one
coming from the Monte Carlo weight assignment, the ``Optimal'' one
coming from the LP formulation in this section.

The table in Fig.~\ref{fig:best1} shows that there exist some weight
assignments that can guarantee the top rank to 27 universities. Weight
assignments also exist for moving the next five universities to the
rank 2 position, for moving the next six universities to the rank 3
position, and so on.

Both of these tables also show the key difference between the Monte
Carlo search and the optimal search to find the top rank. By
optimality and also as these tables demonstrate, the ``Optimal'' ranks
necessarily lower than or equal to the ``Random'' ranks; however, for
some universities the differences are quite large. Recall that the
Monte Carlo search used 10,000 runs while the optimal search used a
single run. On the other hand, the Monte Carlo search finds the top
ranks for every university while the optimal search needs a new run to
find the top rank for each university. Despite these differences, the
optimal search is far faster to run for 10s of universities.

What is the implication of the experiments in this section? Recall
that any of the university rankings assigns the top rank to a single
university. The experiments in this section indicate that actually 27
universities can attain the top rank under some weight
assignments. Does it then make sense to claim that only a single
university is the top university? This section indicates that the
answer has to be a no. Then, how can we rank universities based on the
results of this section? We think within the limits of this section,
we may rank universities in groups, the top group containing the
universities that can attain rank 1, the next group containing the
universities that can attain rank 2, and so on.

One counter-argument to the argument in the paragraph above may be
the realization that universities attain the top rank under some but
different weight assignments. In other words, a given weight
assignment that moves a particular university to its best rank may not
make another university to attain its best rank. This means a single
ranking cannot be used to rank the universities, which is against the
idea of university rankings in the first place. At the same time a
single ranking does not give the correct picture to the interested
parties, e.g., the students trying to choose a university to apply to.

\begin{figure}[ht]
  \centering
  \includegraphics[scale=0.3]{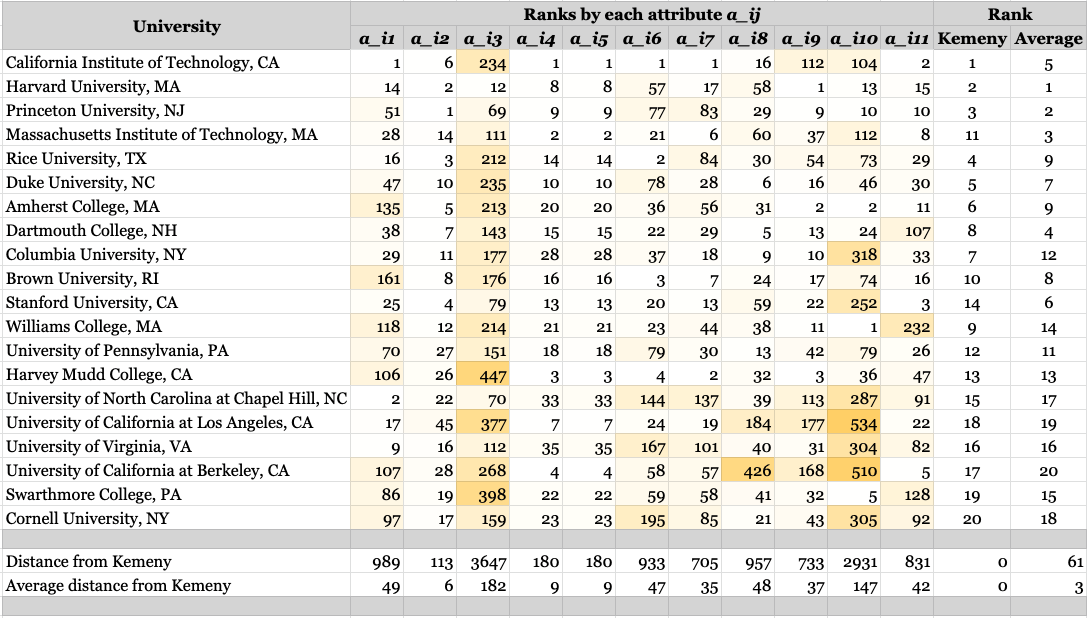}
  \caption{The rank of each university with respect to each attribute
    for the universities in our original ranking. Here the label {\em
      a\_ij} refers to $a_{ij}$.}
  \label{fig:top20kemeny1}
\end{figure}

\begin{figure}[ht]
  \centering
  \includegraphics[scale=0.3]{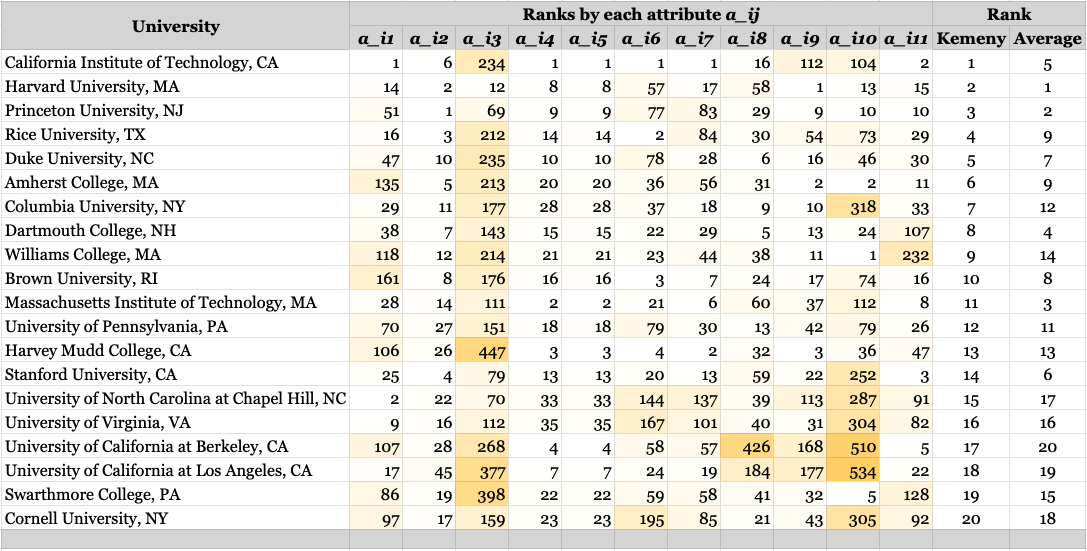}
  \caption{Our original ranking reranked with respect to Kemeny
    ranks. Here the label {\em a\_ij} refers to $a_{ij}$.}
  \label{fig:top20kemeny2}
\end{figure}

\begin{figure}[ht]
  \centering
  \includegraphics[scale=0.3]{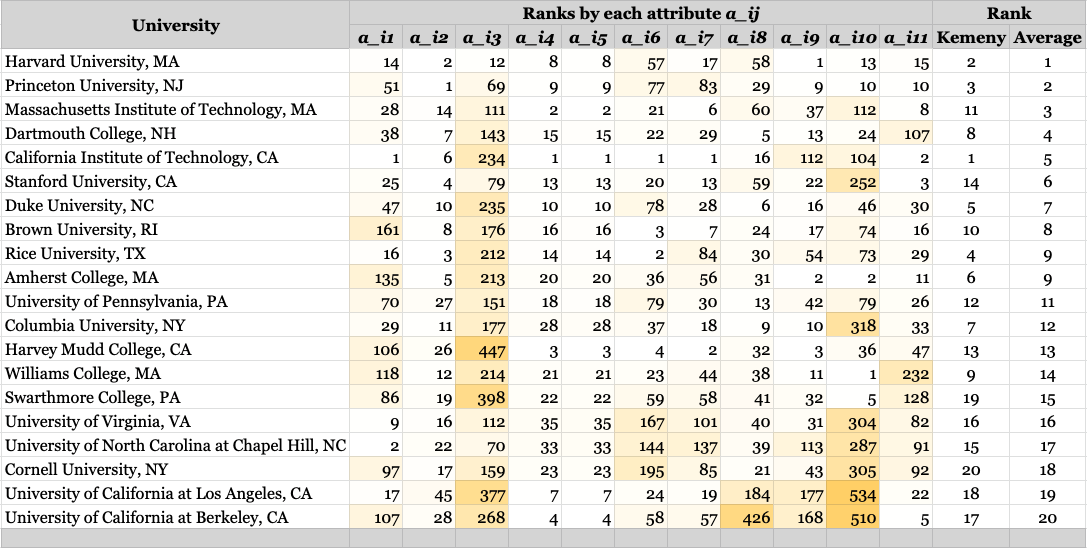}
  \caption{Our original ranking reranked with respect to average ranks
    over all the attribute-based ranks. Here the label {\em a\_ij}
    refers to $a_{ij}$.}
  \label{fig:top20kemeny3}
\end{figure}

\subsection{Problem 5: Rankings without Weights}
\label{sec:kemeny}

In every problem we solved above, we had to derive weights. In this
section, we will explore the possibility of rankings that do not use
weights at all. The idea is to rank every university for each
attribute independently, and then aggregate these rankings into a
final joint ranking. This ensures that if the attributes themselves
(apart from which ones are selected) are objective enough, the
per-attribute ranking is also objective. This leads to a far more
objective ranking that any weight-based rankings. Similar proposals,
independently proposed, are in \cite{At2015,Ro2013}.

The problem of aggregating multiple independent rankings into a final
joint ranking is called the ``rank aggregation'' problem in the
literature~\cite{DaDa2011,DwKuNa2001}, where the best method to use
depends on the application area. Here we will use the Kemeny rule,
which is recognized as one of the best overall~\cite{DwKuNa2001}.

The Kemeny rule minimizes the total number of disagreements between
the final aggregate ranking (called the Kemeny ranking) and the input
rankings, which are the independent per-attribute
rankings. Unfortunately computing the optimal Kemeny ranking is
NP-hard~\cite{DwKuNa2001}, which means we can either resort to
approximation or heuristic algorithms or we can still seek the optimal
by reducing the problem size. We will go for the latter.

The Kemeny ranking of a set of universities (or objects) can be
computed optimally by solving the following integer linear programming
(ILP) formulation:
\begin{equation}
  \label{eq:kemeny1}
  \begin{array}{ll}
  \mbox{minimize } & \sum\limits_{a\neq b} n_{ba}x_{ab}\\
  \mbox{subject to } & \\
  x_{ab}+x_{ba}=1 & (\forall a, b : a\neq b) \\
  x_{ab}+x_{bc}+x_{ca} \leq 2 & (\forall a, b, c : a\neq b, b\neq c,
  c\neq a) \\
  x_{ab} \in {0,1} & (\forall a, b : a\neq b) \\
  \end{array}
\end{equation}
where for two universities $a$ and $b$, $x_{ab}=1$ if $a$ is ranked ahead
of $b$ in the aggregate ranking or $0$ otherwise, and $n_{ba}$ is the
number of input rankings that rank $b$ ahead of $a$. The second
constraint above can also be written as $x_{ab}+x_{bc}+x_{ca} \geq 1$.

Fig.~\ref{fig:top20kemeny1} and Fig.~\ref{fig:top20kemeny2} show the
result of the Kemeny ranking. In both figures, we have the rank of
each university $i$ per attribute $a_{ij}$. The ranks were input to
the integer linear program in Eq.~\ref{eq:kemeny1}. The ``Kemeny
rank'' column shows the resulting ranks for each university. For
comparison, the ``Average rank'' column shows the average rank over
all the per-attribute ranks for each university. Here the difference
between the former and latter figure is that the former shows the
universities in our original top20 ranking as in
Fig.~\ref{fig:top20ours} whereas the latter reranks this top 20 based
on the derived Kemeny ranks.

In Fig.~\ref{fig:top20kemeny1}, the last row gives the average
similarity score between each per-attribute ranking and the Kemeny
ranking (computed using the Spearman's footrule distance). A close
inspection of the per-attribute (col) ranks and the similarity score
reveals two interesting observations: a) the ranks based on $a_{i3}$
and $a_{i10}$ are very large; b) the ranks for $a_{i2}$, $a_{i4}$, and
$a_{i5}$ are usually small, with the first one being the smallest. The
``large'' and ``small'' designations also apply the similarity
score. It may be possible to reason from these observations that those
attributes that produce a ranking too dissimilar to the Kemeny ranking
may not be good attributes to use for university ranking but we will
leave this as a conjecture for future research at this point.

You may wonder whether or not it is possible (as done in Problem 2) to
find a set of weights to guarantee the final Kemeny ranking for the
top 20. The answer to this is unfortunately negative unless NP=P. The
reason is that Problem 2 can be solved in polynomial time whereas
Problem 3 is NP-hard; unless NP=P, we cannot use a polynomial time
algorithm to solve an NP-hard problem. This means that a Problem 2
version of the Kemeny ranking for the top 20 is necessarily an
infeasible linear program. There are techniques, e.g., via the use of
slack variables as shown in \cite{lpinfeasible2019}, to discover a set
of weights that add up to 99\% instead of 100\% as required but a full
exploration of this avenue is left for future research.

Finally, if the number of universities is too many, it is possible to
use approximate algorithms for Kemeny ranking. If all else fails, even
using the average rank over the per-attribute as a crude approximation
to Kemeny ranking may work. For example, the similarity distance
between the average and Kemeny rankings is 3, the best over all the
per-attribute rankings. Note that average ranks can be found quickly
in polynomial time.

\begin{figure}[ht]
  \centering
  \subfloat[Histogram (pdf)]{{\includegraphics[scale=0.3]{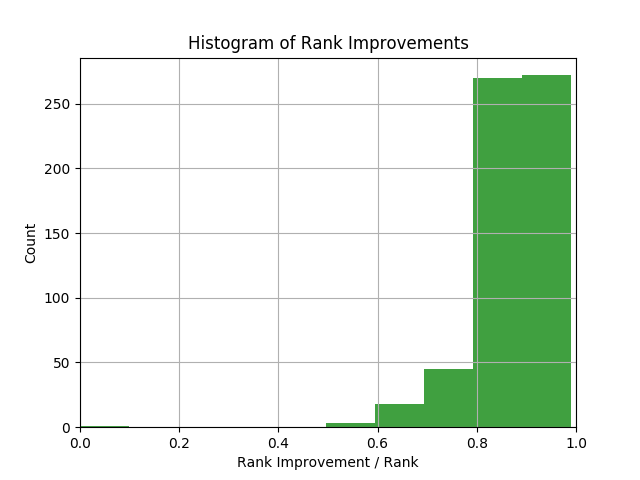}}}
  \qquad
  \subfloat[Cumulative histogram (cdf)]{{\includegraphics[scale=0.3]{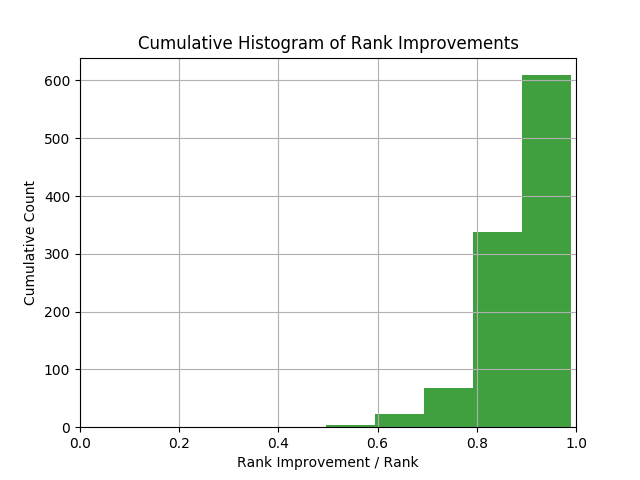}}}  
  \caption{The histogram (left) and cumulative histogram (right) of
    the percentage of the maximum rank improvement of a university
    with respect to old rank of the university for Case One (one
    university improving at a time). Most of the improvements are
    above 80\%.}
  \label{fig:y2}
\end{figure}

\begin{figure}[ht]
  \centering
  \subfloat[Histogram (pdf)]{{\includegraphics[scale=0.3]{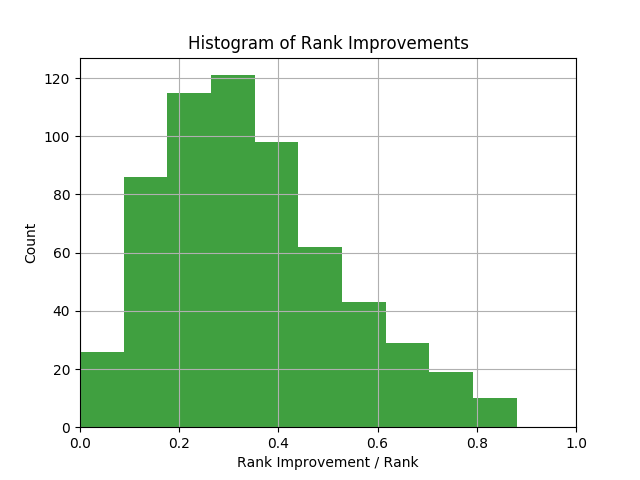}}}
  \qquad
  \subfloat[Cumulative histogram (cdf)]{{\includegraphics[scale=0.3]{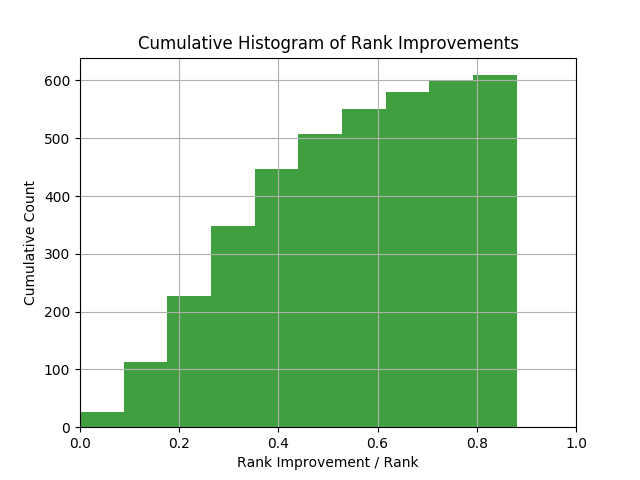}}}  
  \caption{The histogram (left) and cumulative histogram (right) the
    percentage of the maximum rank improvement of a university with
    respect to old rank of the university for Case All (all
    universities improving simultaneously). A good number of
    improvements are in the vicinity of 20\% to 40\%.}
  \label{fig:y1}
\end{figure}

\subsection{Problem 6: Improving Rankings}
\label{sec:delta}

The problems we studied so far have shown that there are many ways of
creating reasonable rankings. One problem left is the question of
which attributes a university should focus on to improve its
ranking. This section proposes a simple solution to this problem.


One question to address is how many attributes a university should
improve at the same time. The extreme answer is all of the attributes
but this is probably unrealistic due to the large amount of resources
such a focus would require. For simplicity, we will assume that the
university focuses on only one attribute, the one that provides the
best improvement in the score of the university.

Another question is how many other universities are improving their
scores at the same time that the university in question is focusing on
its own. The answer is difficult to know but probably the answer is
most of the universities due to the drastic impact, positive or
negative, of the university rankings. Again for simplicity, we will
consider one attribute at a time with the following two extremes:
Given an attribute, one extreme is only the university in question
modifies the value of the attribute (called ``Case One''), and the
other extreme is every university modifies the value of the attribute
simultaneously (called ``Case all''). For the university in question,
the improvement in score is likely somewhere in between of these
extremes.

Note that we will not delve into what it takes in terms of resources,
e.g., time, money, staff, etc., for a university to improve its
score. The cost of these resources is expected to be
substantial. Interested readers can refer to the relevant references,
e.g., \cite{Ku2014}.

Let us start with some definitions. Let $a_{ij}^{*}$ denote the
maximum value of the attribute $a_{ij}$ for the $i$th
university. Recall that due to the attribute value normalization, this
maximum value is at most 1.0. Also recall that in our formulation,
larger values of each attribute are the desired direction to maximize
the scores.

Earlier in Eq.~\ref{eq:score} we defined the score $s_{i}$ of the
$i$th university in the ranking as
\begin{equation}
  s_{i}^{old} = \sum\limits_{j=1}^{m} w_{j} a_{ij},
\end{equation}
which we will refer to as the $old$ score due to the change we will
introduce to compute the $new$ version. Suppose we maximized the value
of the $k$th attribute; then the new score becomes
\begin{equation}
  s_{i}^{new} = \sum\limits_{j=1}^{m} w_{j} a_{ij} - w_{k}a_{ik} + w_{k}a_{ik}^{*},
\end{equation}
and the improvement in the score
\begin{equation}
  \Delta s_{i} = s_{i}^{new} - s_{i}^{old} =  w_{k} (a_{ik}^{*} - a_{ik}),  
\end{equation}
which is guaranteed to be non-negative.

Our algorithm for this problem follows the following steps: 1) find
the maximum of each attribute across all universities in the input
list of universities; 2) compute new scores for each pair of
university and attribute; 3) sort the list of universities using the
new scores; 4) print new ranks and rank changes. In step 3, sorting is
done over each attribute and for both cases, Case One and Case All.

The results are given using histograms in Fig.~\ref{fig:y2} for Case
One and in Fig.~\ref{fig:y1} for Case All. In each plot in these
figures, the x-axes represent the percentage of the maximum rank
improvement, i.e., $\Delta s_{i} / s_{i}^{old}$, in ten buckets; the
y-axes represent different things: The y-axis on the left is the count
or number of universities falling into each bucket on the x-axis
whereas the y-axis on the right is the cumulative count or number of
the universities falling into each bucket from the one on the left up
to the one corresponding to the count.

These histograms show that drastic rank improvements are possible for
Case One, probably as expected due to the changes happening one at a
time. For the majority of universities, the rank improvements can be
above 80\%. As for Case All, the rank improvements are still
impressive, half of them in the vicinity of 20\% to 40\%.

\begin{figure}[ht]
  \centering
  \includegraphics[scale=0.4]{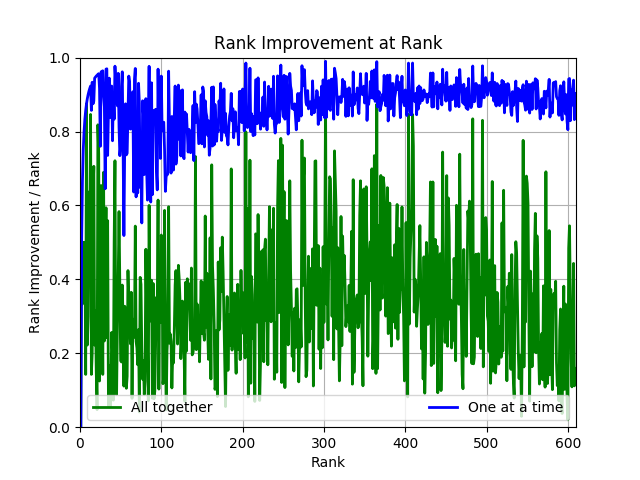}
  \caption{Percentage of the maximum rank improvement of a university
    with respect to the old rank of the university as a result of
    attribute value improvements. The blue or upper plot is for Case
    One (one university improving at a time) and the green or lower
    plot is for Case All (all universities improving
    simultaneously). These plots indicate a good amount of consistency
    in rank improvements across ranks.}
  \label{fig:allone}
\end{figure}

Fig.~\ref{fig:allone} shows how the percentage of the maximum rank
improvement changes with respect to the old rank for both Case One
(blue or upper plot) and Case All (green or lower plot). These plots
show that the rank improvements are roughly consistent across
ranks. This also means that for lower ranked universities, rank
improvements can be significant. Note that the two figures above are
distribution or histogram versions of the data in this figure.

\begin{figure}[ht]
  \centering
  \includegraphics[scale=0.3]{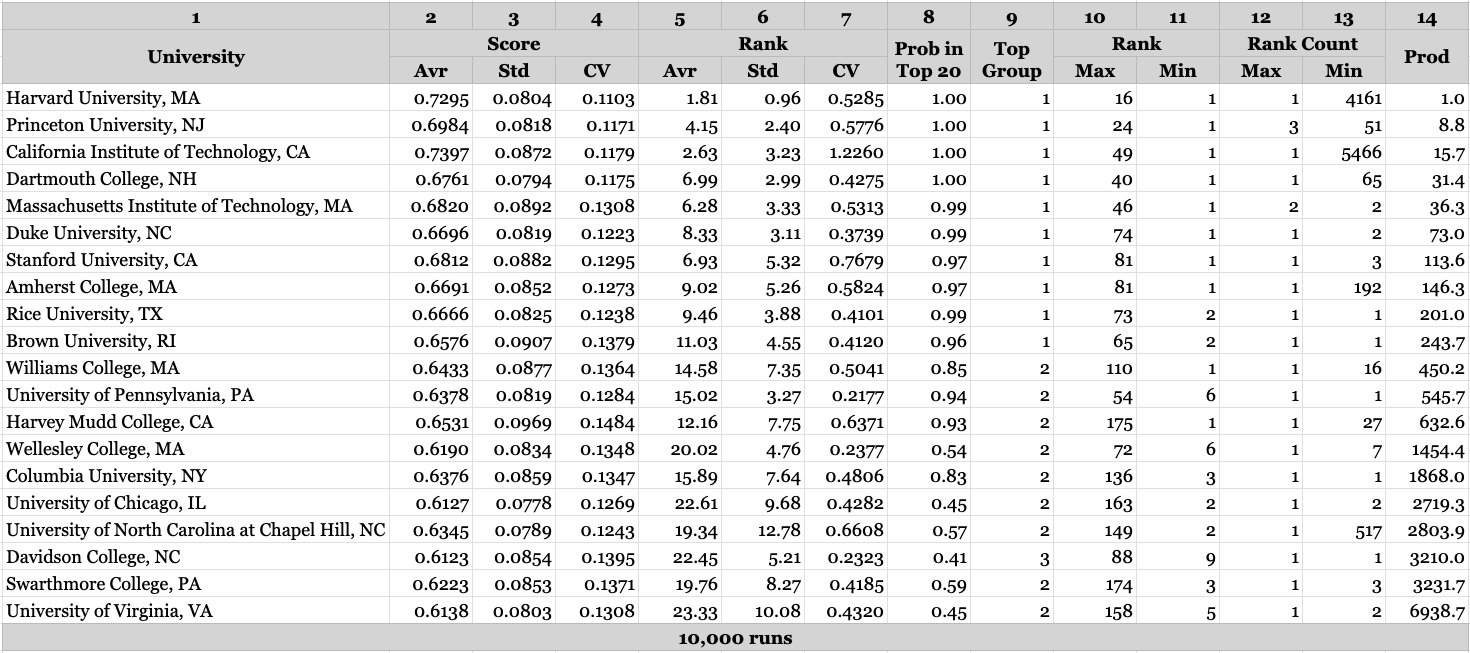}
  \caption{Ranking of the top 20 universities in increasing ``Prod''
    (Column 14) order using the arithmetic mean formula and using
    uniformly random weights in all 10,000 runs. ``Avr'', ``Std'', and
    ``CV'' in Columns 2-7 stand for the average, the standard
    deviation, and the coefficient of variation, respectively. ``Prob
    in Top 20'' in Column 8 is the probability of falling in the top
    20 universities when they are ranked in decreasing score
    order. Each column is explained in \S~\ref{sec:random}.}
  \label{fig:random-arith-p}
\end{figure}

\begin{figure}[ht]
  \centering
  \includegraphics[scale=0.3]{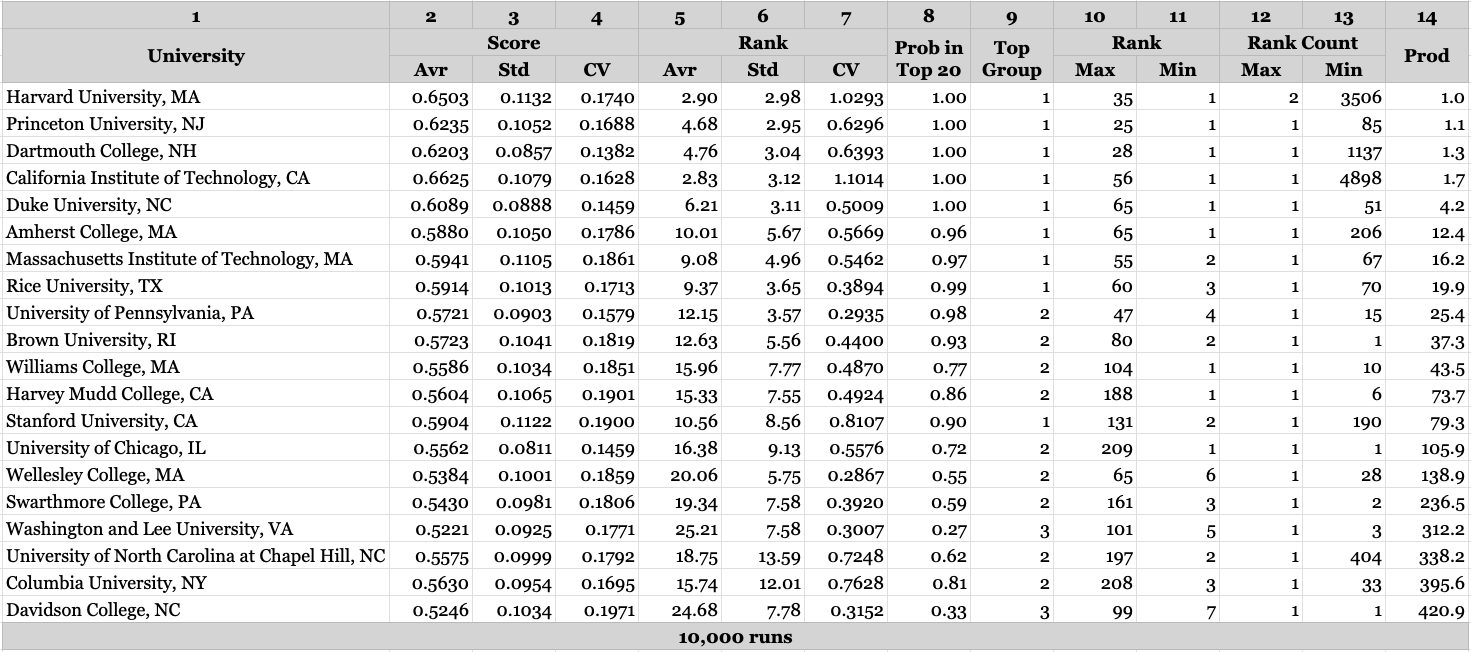}
  \caption{Ranking of the top 20 universities in increasing ``Prod''
    (Column 14) order using the geometric mean formula and using
    uniformly random weights in all 10,000 runs. ``Avr'', ``Std'', and
    ``CV'' in Columns 2-7 stand for the average, the standard
    deviation, and the coefficient of variation, respectively. ``Prob
    in Top 20'' in Column 8 is the probability of falling in the top
    20 universities when they are ranked in decreasing score
    order. Each column is explained in \S~\ref{sec:random}.}
  \label{fig:random-geo-p}
\end{figure}

\section{Discussion and Recommendations}
\label{sec:reco}

Let us summarize the different ways we can produce a ranking. We first
need to select attributes. The universities in our dataset have 52
attributes, 50 of which are suitable as attributes. Let us assume that
we wanted to select 20 attributes for our ranking. The number of ways
of selecting 20 attributes out of 50 attributes is approximately
$4.7\times 10^{13}$, i.e., roughly 47 trillion!

Then we need to decide on whether or not to use weights at all. If we
decide not to use any weights, then we have to use one of the rank
aggregation algorithms. Each of these algorithms is highly likely to
produce a different ranking.

If we decide to use weights, then we have multiple choices to select
them: Uniform weights, non-uniform weights derived subjectively,
non-uniform weights derived randomly, non-uniform weights derived
optimally (which in turn has multiple possibilities based on the
objective function used).

Together with the weights, we also need to select which aggregation
formula to use: Arithmetic or geometric. The combination of how to
derive weights and how to aggregate them lead to different rankings.

Once these are selected, we next need to decide on whether or not we
will derive the best rank each university can attain or run Monte
Carlo simulation. The latter leads to more ways of ranking
universities based on one of these factors: The average score, average
rank, etc., many of the columns in Fig.~\ref{fig:random-arith-p} and
Fig.~\ref{fig:random-geo-p}.

All in all, the discussion above shows that there are many ways of
ranking universities, each way with its own pros and cons. Hope this
may provide more evidence on the reliability of university rankings.

At this point a good question is what we would recommend. First, we
would like to clarify that our aim in this paper is not to present a
better way of ranking universities; we hope to come to realizing this
aim in a future study. However, we will mention a couple of ways that
might be interesting to explore further.

Before we start, we would like to emphasize that we support the Berlin
Principles~\cite{Berlin2019}. These principles provide good guidelines
for ranking higher education institutions, which include
universities. We will assume that the attributes for ranking will be
selected according to these guidelines. In accordance with these
principles, we also open source our datasets and software
code~\cite{Da2020}.

Regarding our recommendations, we prefer ranking universities without
weights as this paper and many in the literature convincingly show
that weight-based rankings are not very reliable. In this regard, a
rank aggregation method like the Kemeny rule is a good choice.

If weights are to be preferred, we do not recommend the use of
subjective weights nor any subjective attributes. Subjective
attributes make it difficult to replicate the ranking while subjective
weights are open to gaming, as shown in this paper and in the
literature. What we recommend is to present universities in groups
rather than in a forced rank linear order. The group boundaries can be
determined using either the ILP in \S~\ref{sec:best} (``the best
possible rank'' case) or the Monte Carlo simulation in
\S~\ref{sec:random} (the ``Monte Carlo'' case) or potentially using
both. Note that both of these methods use weights as an intermediate
mechanism to reach their conclusions.

In the best possible rank case we have groups of universities where
groups are ranked but not the universities inside these groups. We say
a group for rank $i$ may consist of all universities whose best
possible rank is $i$. For example, the top group can consist of all
the universities that are proven to attain rank 1, as in
Fig.~\ref{fig:best1}; the next group is for those that can attain rank
2 at best, and so on, as in Fig.~\ref{fig:best2}. Note that in this
case it is very likely that the group sizes will not be the same.

A disadvantage of the best possible rank case is that a university may
attain a top rank but its probability of occurrence may be tiny. This
may be alleviated using the Monte Carlo case. In the Monte Carlo case,
two good things are happening: One is that the average score ranking
converges to the uniform-weight case, and the other is that a
simulation of a huge number of weight assignments gets performed,
potentially subsuming many of the weight assignments that may be
performed by committees or users of the ranking such as students or
parents. Moreover, we can collect many statistics as shown in
\S~\ref{sec:random}.

In the Monte Carlo case, we may assign universities to groups of a
certain size, e.g., 10, in that the top group is for the ranks from 1
to 10, the second best group is for the ranks from 11 to 20, and so
on. A university may be assigned to a group for the ranks from $i$ to
$i+9$ if the university attains the ranks for this group more often
than the other ranks, i.e., ranks smaller than $i$ or larger than
$i+9$. If there are ties for a university, we may assign the
university to the highest rank group. In Fig.~\ref{fig:random-arith-p}
and Fig.~\ref{fig:random-geo-p}, Column 9 gives the group ids
according to this way of assigning groups.

This way of group assignment is actually not a good way if the
rankings are done in score orderings. This is because of the high
correlation between these group assignments and the score ordering,
which follows by the first observation made in \S~\ref{sec:random}. In
other words, we need to find another way of determining group
assignments.

We conclude this section by sharing a new heuristic way of creating a
ranking, as shown in Fig.~\ref{fig:random-arith-p} and
Fig.~\ref{fig:random-geo-p}. This heuristic uses a product,
unsurprisingly called ``Product'' in Column 14 of these figures, of
four values computed in the Monte Carlo simulations: The average of
ranks, the standard deviation of ranks, the minimum rank, and the
difference between the maximum and minimum ranks. The heuristic states
that the smaller any of these values is, the better the corresponding
university is. By multiplying these values, we magnify this effect in
that the smaller the product of these four values, the better the
corresponding university is. In the figures, we give the product as a
relative value by diving each product by the smallest product. Note
that according to this product, the top two universities turn out to
be Harvard University and Princeton University, in both the arithmetic
and geometric cases.

\section{Conclusions}

Rankings of universities are attention-grabbing events in the public
due to their impact on students, parents, universities, funding
agencies, and even countries. Among a large number of such rankings,
four are well known and attract the most interest.

These rankings use a similar methodology that ranks universities based
on their scores, usually computed as a sum-of-products formula
involving a set of attributes and their respective weights, all
subjectively selected by the rankings organizations. There is a huge
literature on these rankings and many issues of theirs.

In this paper, we produce a university ranking of our own in a
repeatable way and in the open by applying a generic ten-step ranking
methodology to a public dataset of US universities. Using formal and
algorithmic formulations on this ranking as our testbed, we explore
multiple problems and provide convincing evidence that university
rankings as commonly done today using the same generic ranking
methodology (though different attributes and weights) are not reliable
in that it is relatively easy to move many universities to the top
rank or automatically generate many reasonable rankings with appealing
weights. Given the many applications of the generic ranking
methodology in ranking objects other than universities, we believe our
findings have wide applicability.

We share our datasets and software code in a public
repository~\cite{Da2020} to ensure repeatability and encourage further
studies.

\bibliographystyle{plain}

\end{document}